\documentclass{elsart}

\usepackage{epsfig}
\usepackage{amssymb}
\usepackage{amsmath}

\newcommand{\pre}{Physical Review E}
\newcommand{\be}{\begin{equation}}
\newcommand{\ee}{\end{equation}}
\newcommand{\bea}{\begin{eqnarray}}
\newcommand{\eea}{\end{eqnarray}}
\begin{document}

\begin{frontmatter}

\title{Balanced vehicular traffic at a bottleneck}
\author{Florian Siebel and Wolfram Mauser}
\ead{\{f.siebel, w.mauser\}@iggf.geo.uni-muenchen.de}

\address{Department of Earth and Environmental Sciences, University of
  Munich, Luisenstra\ss e 37, D-80333 Munich, Germany}
\author{Salissou Moutari and Michel Rascle}
\ead{\{salissou, rascle\}@math.unice.fr}

\address{Laboratoire J. A. Dieudonn\'e, UMR CNRS N$^\circ$ 6621, 
Universit\'e de \\Nice-Sophia Antipolis, Parc Valrose, 06108 Nice Cedex 2, 
France}

\begin{abstract}
The balanced vehicular traffic model is a macroscopic model for vehicular
traffic flow. We use this model to study the traffic dynamics at highway 
bottlenecks either caused by the restriction of the number of lanes or by
on-ramps or off-ramps. The coupling conditions for the Riemann problem of 
the system are applied in order to treat the interface between different 
road sections consistently. Our numerical simulations show the appearance 
of synchronized flow at highway bottlenecks.
\end{abstract}

\end{frontmatter}

\section{Introduction}
\label{intro}
The balanced vehicular traffic model (BVT model), which was first introduced
in~\cite{SiM205}, generalizes the macroscopic traffic model of Aw, 
Rascle~\cite{AwR00,Zha02} and Greenberg~\cite{Gre01} by introducing an 
effective relaxation coefficient into the momentum equation of traffic flow. 
This effective relaxation coefficient can become negative, resulting in
multivalued fundamental diagrams in the congested regime. Such negative 
effective relaxation coefficients follow from finite reaction and relaxation 
times of drivers (see~\cite{SiM205}). For related ideas, see 
Greenberg-Klar-Rascle~\cite{GKR}, Greenberg~\cite{Gre04}.

In a previous work~\cite{SiM305} we studied the behavior of the BVT model 
at a bottleneck. There, we manipulated the partial differential equations
describing traffic flow at the bottleneck by artificially resetting the 
average velocity in order to model a speed restriction. The results 
obtained in that study showed the basic behavior of the BVT model and 
its potential to explain the observed patterns of traffic flow~\cite{Ker04}. 
Although the procedure of resetting quantities,
i.e. the average velocity in that case, was locally restricted, it leaves 
the question whether the model can adequately describe synchronized flow at a 
bottleneck. In order to show this, the current paper systematically studies
the traffic dynamics at highway bottlenecks in the BVT model by numerical 
means. 
To do this in agreement with the underlying partial differential equations we 
use the coupling conditions for the Riemann problems at the interface between 
different highway sections.  
We focus the discussion on two setups: In the first setup we study a
bottleneck caused by the narrowing of a highway from three lanes to two
lanes. In the second setup we study a two-lane highway with an on-ramp and
an off-ramp.

The outline of the paper is as follows. In Section~\ref{coupling} we
summarize the theory of the coupling conditions for the Riemann problem at
intersections. Whereas Section~\ref{lanerestrict} presents numerical results 
for a highway bottleneck caused by the restriction of the number of lanes 
from three lanes to two lanes, Section~\ref{onofframp} presents the numerical 
results for a two-lane highway with an on-ramp and an off-ramp. In 
Section~\ref{extension} we introduce the possible generalization to an 
arbitrary junction. We finally summarize our results in 
Section~\ref{conclusion}.

\section{Coupling conditions}\label{coupling}
Before we have a closer look at the coupling conditions, let us repeat the
principal equations of the BVT model. The evolution equations for the density
$\rho$ of vehicles and the average velocity $v$ are described by the following
hyperbolic system of balance laws
\bea
\label{rhob}
\frac{\partial \rho}{\partial t} + \frac{\partial (\rho v)}{\partial x} & = &
0, \\
\frac{\partial (\rho(v-u(\rho))}{\partial t} + \frac {\partial (\rho v
  (v-u(\rho)))}{\partial x} & = & b(\rho,v) {\rho (u(\rho)-v)}.
\label{vb}
\eea
Here, $u(\rho)$ denotes the equilibrium velocity, therefore the expression 
$v-u(\rho) = w$ describes a distance to equilibrium. The quantity 
$b(\rho,v)$ is the
effective relaxation coefficient. For the case where $b(\rho,v)$
becomes negative, there are additional equilibrium velocities, i.e. high-flow
branch and the jam line. Whereas the high-flow branch is metastable for
intermediate densities and unstable for high densities, the jam line is
unstable for intermediate densities and metastable for high densities. For
a detailed discussion see~\cite{SiM305}. Note that the pseudo-momentum
$\rho(v-u(\rho))$ is not conserved due to the non-vanishing term on the
right-hand side of Eqn.~(\ref{vb}). This source term plays an essential
role for the traffic dynamics on road sections, but it is neglected for 
the situation where one is interested in the Riemann problems at 
intersections, since it is never a delta-function.

\subsection{Background}
Piccoli and Garavello \cite{GaP06} appear to have been the first to propose an
intersection modeling by using the Aw-Rascle ``second order'' model of traffic 
flow \cite{AwR00}. In their approach, only the mass flux is conserved but not 
the pseudo-momentum. In \cite{HeR06}, Herty and Rascle proposed another 
approach in which mass flux and pseudo-momentum are both conserved. But they 
maximized the mass fluxes at the intersection with some arbitrary {\em given}
homogenization coefficients. 
In \cite{HMR06}, the latter approach has been generalized by maximizing the 
total mass flux at the junction without fixing any condition. In fact, the homogenization 
coefficients are not arbitrary but obtained directly from the mass flux maximization. Another 
approach also based on the mass flux maximization is given in \cite{HaB05}. 
Our approach in the current paper is similar to the latter one, in particular 
we maximize the total flux at the junction, at the same time conserving the 
pseudo-momentum of the original ``Aw-Rascle'' system. Here, we are 
particularly concerned with the BVT model \cite{SiM205}. We note that our  
treatment of the homogenization problem which naturally arises in a 
merge junction is different from the one in \cite{HeR06},\cite{HMR06}, 
see Remark~\ref{rem1}.

\begin{figure}[htpb]
\begin{minipage}[t]{.32\linewidth}
  \centering\epsfig{figure=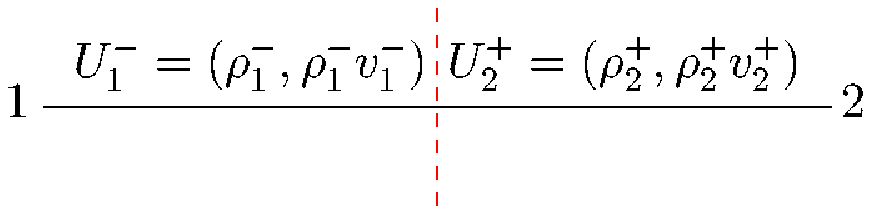,width=\linewidth}
\small a) 
\end{minipage} \hfill
\begin{minipage}[t]{.32\linewidth}
  \centering\epsfig{figure=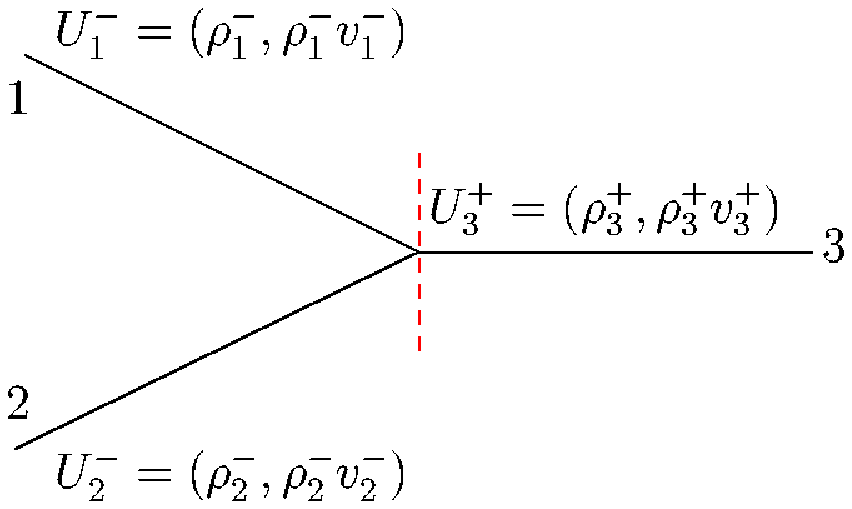,width=\linewidth}
\small b)
\end{minipage} \hfill
\begin{minipage}[t]{.32\linewidth}
  \centering\epsfig{figure=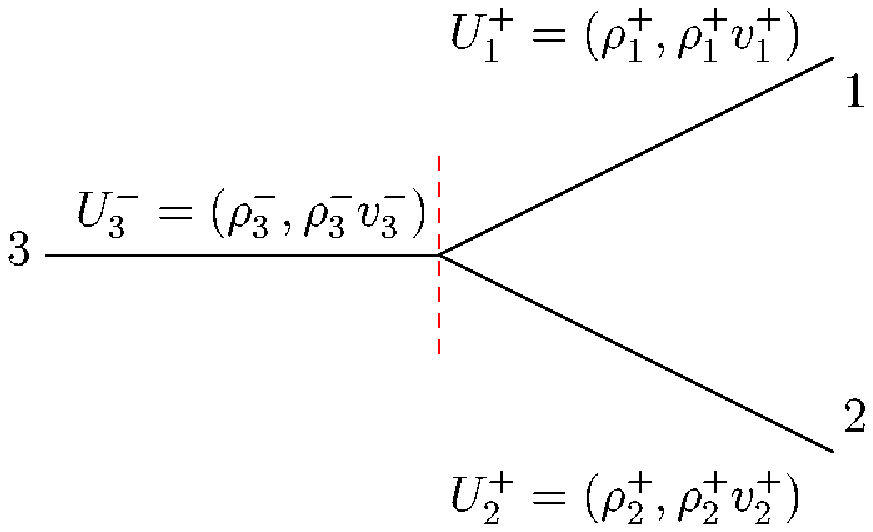,width=\linewidth}
\small c) 
\end{minipage}
\vspace{0.2cm}
\caption{Definition of the quantities at junctions, which are necessary to
  construct the boundary fluxes. The junction in a) consists of one incoming
  and one outgoing road. The panels b) and c) display a merge junction and a
  diverge junction, respectively. The direction of the flow is from left to
  right.}
\label{fig1}
\end{figure}
\subsection{Junction consisting of one incoming and one outgoing road}
In the following we restrict the discussion to the Riemann problem at the
interface between two road sections. The road section 1 is located upstream
and the road section 2 downstream of the interface, where the flux functions 
have to be determined.
Let $U_{1}^{-}=(\rho_{1}^{-}, \rho_{1}^{-} v_{1}^{-})$ be the state vector at 
the interface upstream in the road section 1
and let $U_{2}^{+}=(\rho_{2}^{+}, \rho_{2}^{+} v_{2}^{+})$ be the state vector
 at the interface downstream in the road section 2 (see panel a) of 
Fig.~\ref{fig1}). For each road section $j$ we introduce the function $w_{j}$ 
of the state variable $U=(\rho,\rho v)$
\be
\label{w}
w_{j}(U) =  v - u_{j}(\rho).
\ee
According to Herty and Rascle~\cite{HeR06} the fluxes at the interface between 
the two road sections (see Eqs.~(\ref{rhob})-(\ref{vb}))
\be
\label{interfaceflux}
\hat{f} = q \left( \begin{array}{c} 1 \\ w_1(U_{1}^{-}) \end{array} \right)
\ee
can be calculated from the expression
\be
q = \min(d_1(\rho_{1}^{-}),s_2(\rho_{2}^{\dagger})),
\ee
where the demand and supply functions $d_1(\rho)$ and $s_2(\rho)$ and the 
density $\rho_{2}^{\dagger}$ are defined below. 

The density $\rho_{2}^{\dagger}$ is defined as the intersection point of 
the two curves $v = v_2^+$, $w_2(U) = w_1(U_{1}^{-})$ and can be obtained by
solving the implicit equation
\be
u_2(\rho_{2}^{\dagger}) - v_2^+ + w_1(U_1^-) = 0.
\ee
Let us define the function
\be
\eta_{d1}(\rho) = \rho u_1(\rho) + \rho w_1(U_1^-).
\ee
For a monotonously decreasing, differentiable function $u_1(\rho)$ 
the function $ \eta_{d1}(\rho)$ has a single maximum at location
$\tilde{\rho}_1$, which can be obtained by solving the implicit equation 
\be
\label{tilderho1}
\eta_{d1}'(\tilde{\rho}_1) = \tilde{\rho}_1 u'_1(\tilde{\rho}_1) + 
u_1(\tilde{\rho}_1) + w_1(U_1^-) = 0.
\ee
It is now possible to define the demand function\footnote{Note that 
in~\cite{SiM05} we
  exchanged the terms demand and supply in comparison to the standard
  notation used here. The demand describes the maximum flow that the road
  section 1 can deliver to the road section 2.}  
\be
d_1(\rho) = 
\begin{cases} 
\eta_{d1}(\rho),& \text{if $\rho \le \tilde{\rho}_1$,} \\
\eta_{d1}(\tilde{\rho}_1),& \text{if $\rho > \tilde{\rho}_1$.}
\end{cases}
\ee
Let us define the function
\be
\eta_{s2}(\rho) = \rho u_2(\rho) + \rho w_1(U_1^-).
\ee
For a monotonously decreasing, differentiable function $u_2(\rho)$ 
the function $ \eta_{s2}(\rho)$ has a single maximum at location
$\tilde{\rho}_2$, which is determined by solving the implicit equation 
\be
\label{tilderho2}
\eta_{s2}'(\tilde{\rho}_2) = \tilde{\rho}_2 u'_2(\tilde{\rho}_2) + 
u_2(\tilde{\rho}_2) + w_1(U_1^-) = 0.
\ee
With this function we define the supply function as
\be
s_2(\rho) = 
\begin{cases} 
\eta_{s2}(\tilde{\rho}_2),& \text{if $\rho < \tilde{\rho}_2$,} \\
\eta_{s2}(\rho),& \text{if $\rho \ge \tilde{\rho}_2$.}
\end{cases}
\ee
We numerically solve the implicit Eqs.~(\ref{tilderho1})
and~(\ref{tilderho2}) using the method of nested intervals. With the
expression for the fluxes at the interface between the two road 
sections (see Eqn.~(\ref{interfaceflux})), we obtain the necessary boundary 
values at the interface for the conservative update scheme described in detail 
in~\cite{SiM205}.
\subsection{Merge junction}
We are now interested in the boundary fluxes at a merge junction, see panel b)
of Fig.~\ref{fig1}. In order to define the demand functions on the incoming
road sections $i=1,2$ we first define the functions $\eta_{di}(\rho)$ as
follows,
\be
\label{demand1}
\eta_{di}(\rho) = \rho u_i(\rho) + \rho w_i(U_{i}^-).
\ee
Let us denote the maxima of the corresponding curves as ${\tilde{\rho}}_i$.
We can then define the demand function of the road section $i$ as
\be
\label{demand2}
d_i(\rho) = 
\begin{cases} 
\eta_{di}(\rho),& \text{if $\rho \le \tilde{\rho}_i$,} \\
\eta_{di}(\tilde{\rho}_i),& \text{if $\rho > \tilde{\rho}_i$.}
\end{cases}
\ee
In order to define the supply function in the road section 3, we first 
introduce
quantities $\beta_i$ describing the fraction of cars entering from the road 
section $i$ into the road section 3. When we assume that the incoming fluxes 
passing through the junction are proportional to the incoming demands, we have
\be
\beta_i = \frac{d_i(\rho_i^-)}{\sum_{j=1}^2 d_j(\rho_j^-)}.
\ee
With this definition, we follow~\cite{HaB05} and do not consider these
fractions as part of the optimization problem as in~\cite{HMR06}.
We define the homogenized value $w_3^*$ for the quantity $w$ defined in 
Eqn.~(\ref{w}), i.e.
\be
w_3^* = \sum_{j=1}^{2}\beta_j w_j(U_j^-),
\ee
and with the latter the function
\be
\label{supply1}
\eta_{s3}(\rho) = \rho u_3(\rho) + \rho w_3^*
\ee
which reaches its maximum value at $\tilde{\rho}_3$. 

\begin{rem}\label{rem1}
Here, we assume that 
the velocity on road 3, near the junction, is given by 
$v= u_3 (\rho) + w_3^*$. This is in contrast with 
\cite{HeR06} and \cite{HMR06}, where the mixture of cars from 
both incoming roads 1 and 2 is assumed to produce an {\em homogenized} flow, 
with a nonlinear relation between $\rho$ and 
$v$ which expresses that the cars from roads 1 and 2 microscopically 
share the available space. The resulting mixture rule is more appropriately 
described in Lagrangian (mass) coordinates, but 
is definitely {\em not} in the above form. 

In other words, the assumptions in \cite{HeR06} and \cite{HMR06} are 
incompatible with the (simpler) formula 
(\ref{supply1}), 
which we assume here. Naturally, in practice the difference is not 
necessarily significant. We will come back to this point elsewhere.
\end{rem}  

We are now able to define the supply 
function
\be
\label{supply2}
s_3(\rho) = 
\begin{cases} 
\eta_{s3}(\tilde{\rho}_3),& \text{if $\rho < \tilde{\rho}_3$,} \\
\eta_{s3}(\rho),& \text{if $\rho \ge \tilde{\rho}_3$.}
\end{cases}
\ee
This function will be evaluated below at the intersection point of the two
curves $v = v_3^+$ and $w_3(U) = w_3^*$, which corresponds to a density
$\rho_{3}^{\dagger}$ fulfilling the implicit equation
\be
u_3(\rho_{3}^{\dagger}) - v_3^+ + w_3^* = 0.
\ee
Finally, we define the downstream boundary fluxes in the road section $i$ as
\be
\hat{f}_i^- = q \beta_i \left( \begin{array}{c} 1 \\ w_i(U_{i}^{-}) \end{array}
\right)
\ee
and the upstream boundary fluxes in the road section 3 as
\be
\hat{f}_3^+ = q \left( \begin{array}{c} 1 \\ w_3^* \end{array} \right),
\ee
where
\be
q = \min \Big( \sum_{j=1}^2 d_j(\rho_j^-), s_3(\rho_{3}^{\dagger})\Big).
\ee
Note that the above boundary fluxes are conserved through the intersection, and
are bounded from above by the demands and supplies.
\subsection{Diverge junction}
In this section we study the boundary fluxes at a diverge junction, see panel 
c) of Fig.~\ref{fig1}. Again we set
\be
\eta_{d3}(\rho) = \rho u_3(\rho) + \rho w_3(U_3^-)
\ee
and denote the maximum of that function as $\tilde{\rho}_{3}$. The demand
function in the road section 3 is defined as
\be
d_3(\rho) = 
\begin{cases} 
\eta_{d3}(\rho),& \text{if $\rho \le \tilde{\rho}_3$,} \\
\eta_{d3}(\tilde{\rho}_3),& \text{if $\rho > \tilde{\rho}_3$.}
\end{cases}
\ee
For the definition of the supply function we first have to prescribe the
fractions of cars intending to enter from road section 3 into road section 1 
and road section 2, which we describe with the quantities $\alpha_{31}$ and 
$\alpha_{32}$. We stress that these quantities - in contrast to~\cite{HaB05} -
need not agree with the actual percentages of the flow from road section 3
entering into road section 1 and road section 2, see 
Eqs.~(\ref{f1p})-(\ref{q2}) below. Under the assumption that all cars 
remain in the network, i.e.  $\alpha_{31} + \alpha_{32} = 1$, we can set
\bea
\label{alpha}
\alpha_{31} & = & \alpha,\\
\alpha_{32} & = & (1 - \alpha),
\eea
with $\alpha \in [ 0,1 ] $. For the outgoing roads sections $k=1,2$ we define
\be
\eta_{sk} = \rho u_k(\rho) + \rho w_3(U_3^-)
\ee
and denote the maxima of these functions $\tilde{\rho}_k$. The supply
function on the outgoing road section $k$ then reads
\be
s_k(\rho) = 
\begin{cases} 
\eta_{sk}(\tilde{\rho}_k),& \text{if $\rho < \tilde{\rho}_k$,} \\
\eta_{sk}(\rho),& \text{if $\rho \ge \tilde{\rho}_k$.}
\end{cases}
\ee
We also have to determine the densities $\rho_k^{\dagger}$, at which these 
supply functions are evaluated. These densities are calculated from the 
intersection of the curves $v = v_k^+$ and $w_k(U) = w_3(U_3^-)$, which 
reduces to the implicit equations
\be
u_k(\rho_k^{\dagger}) - v_k^+ + w_3(U_3^-) = 0.
\ee
Finally, we define downstream boundary fluxes of road section 3 as
\be
\hat{f}_3^- = q \left( \begin{array}{c} 1 \\ w_3(U_{3}^{-}) \end{array}
\right)
\ee
and the upstream boundary fluxes in road section 1 as
\be
\label{f1p}
\hat{f}_1^+ = q_1 \left( \begin{array}{c} 1 \\  w_3(U_{3}^{-}) \end{array} \right),
\ee
and in the road section 2 as
\be
\hat{f}_2^+ = q_2 \left( \begin{array}{c} 1 \\  w_3(U_{3}^{-}) \end{array} \right),
\ee
where
\be
q_1 = \min{\Big(\alpha d_3(\rho_3^-),s_1(\rho_1^{\dagger}) \Big)},
\ee
\be
\label{q2}
q_2 = \min{\Big((1-\alpha) d_3(\rho_3^-),s_2(\rho_2^{\dagger}) \Big)},
\ee
and
\be
\label{qdiverge}
q = q_1 + q_2.
\ee
Note again that the above boundary fluxes are conserved through the interface, 
and are bounded from above by the demands and supplies.
\section{Lane reduction on a highway}
\label{lanerestrict}
\begin{figure}[htpb]
\vspace{0.5cm}
\centering\epsfig{figure=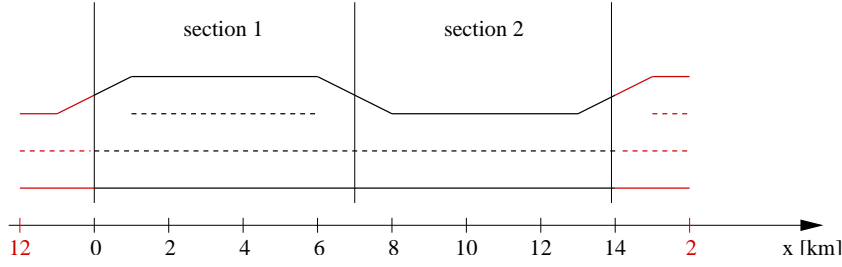,width=0.8\linewidth}
\caption{Sketch of the highway under study. The highway consists of two
road sections of 7 km length. The road section 1 consists of three lanes 
whereas the road
section 2 consists of two lanes. We use periodic boundary conditions,
i.e. the road section 1 is also located downstream of the road section 2.\label{fig2}}
\end{figure}

\begin{figure}[h!]
\vspace{-0.1cm}
\begin{minipage}[t]{.45\linewidth}
  \centering\epsfig{figure=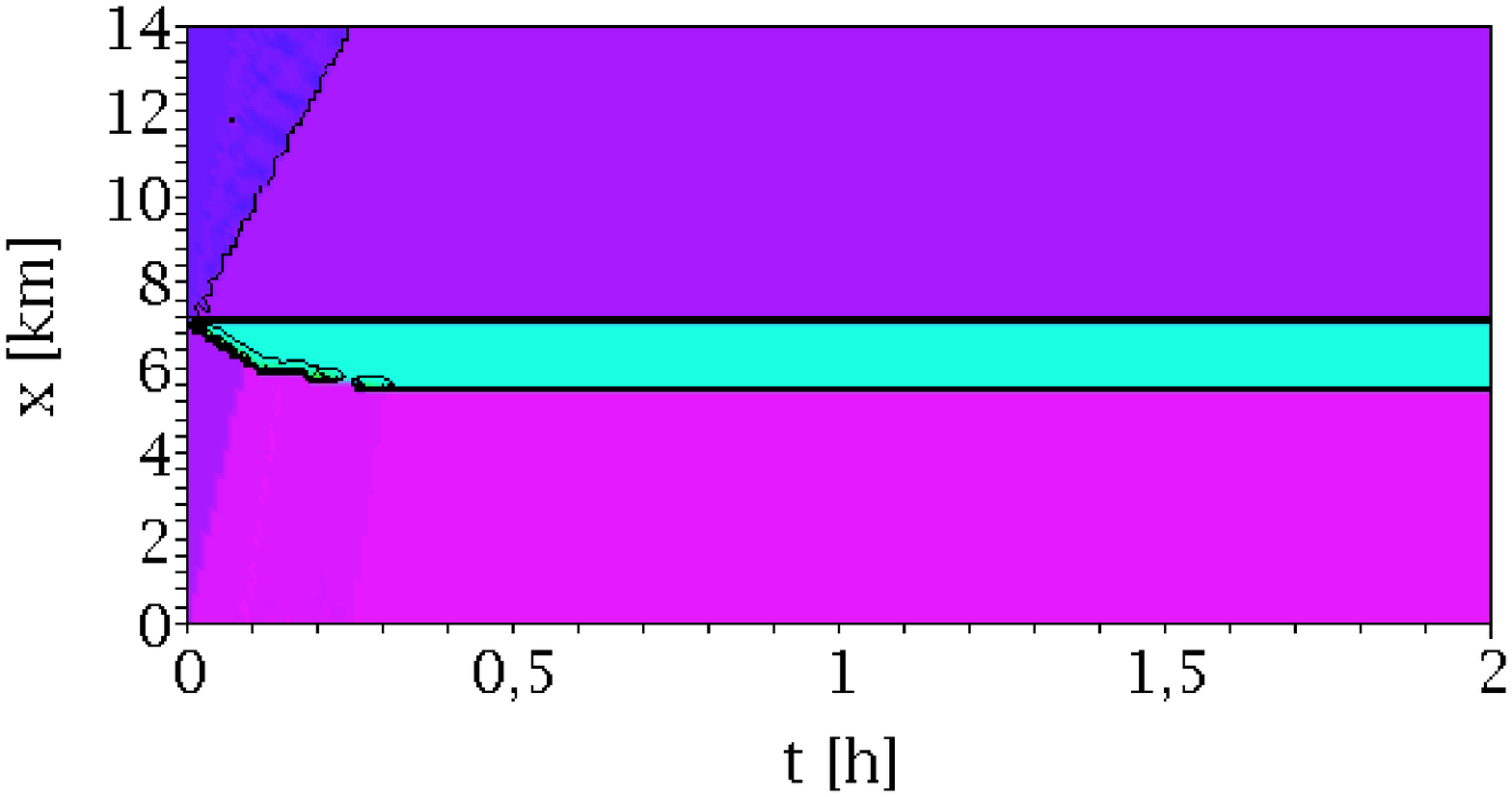,width=\linewidth} 
\end{minipage} \hfill
\begin{minipage}[b]{0.08\linewidth}
\centering \tiny $\rho_0=50~$ [1/km]
\vspace{1.3cm}
\end{minipage} \hfill
\begin{minipage}[t]{0.45\linewidth}
 \centering\epsfig{figure=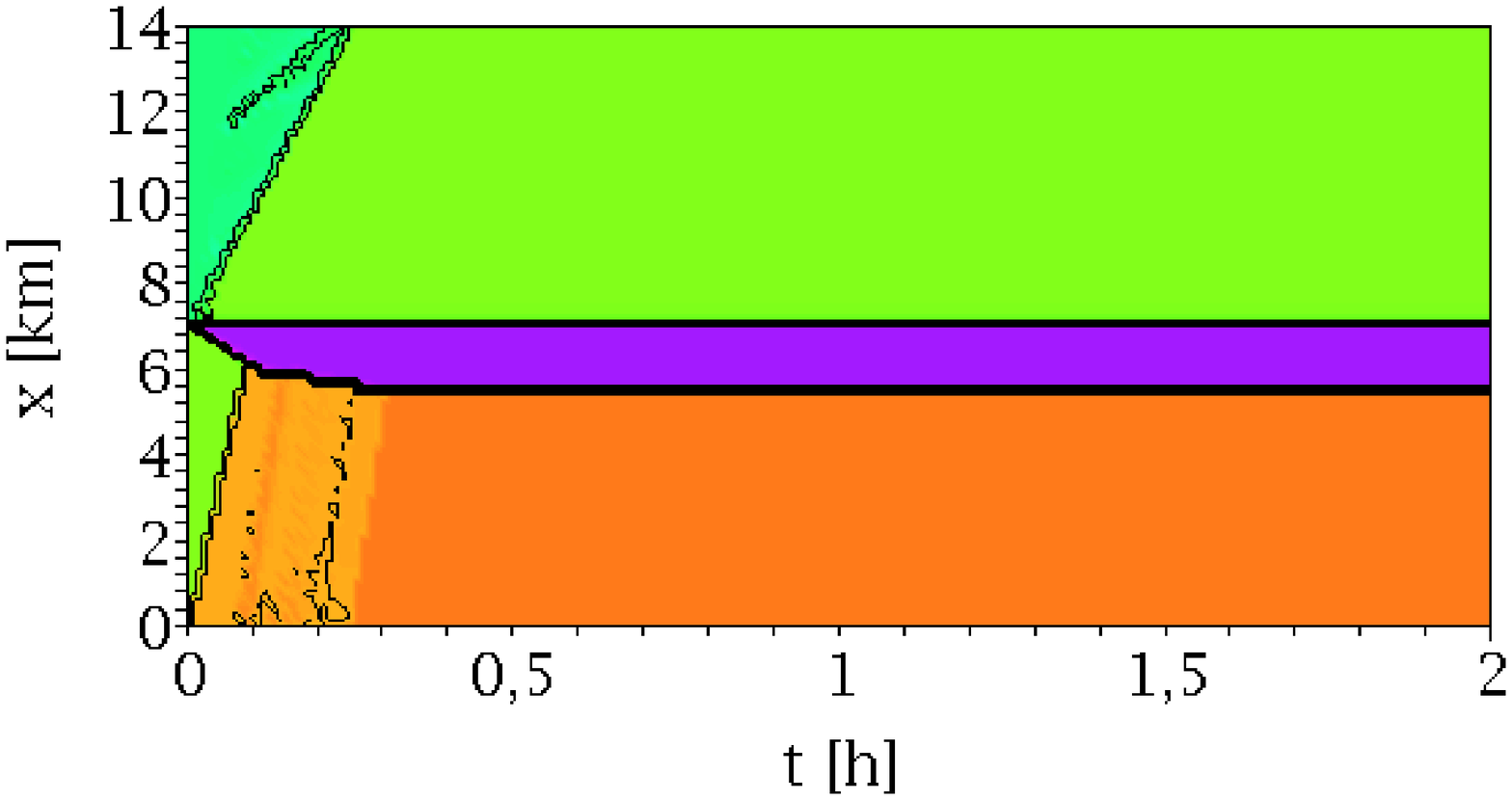,width=\linewidth}
\end{minipage}
\vspace{-0.2cm}
\begin{minipage}[t]{0.45\linewidth}
  \centering\epsfig{figure=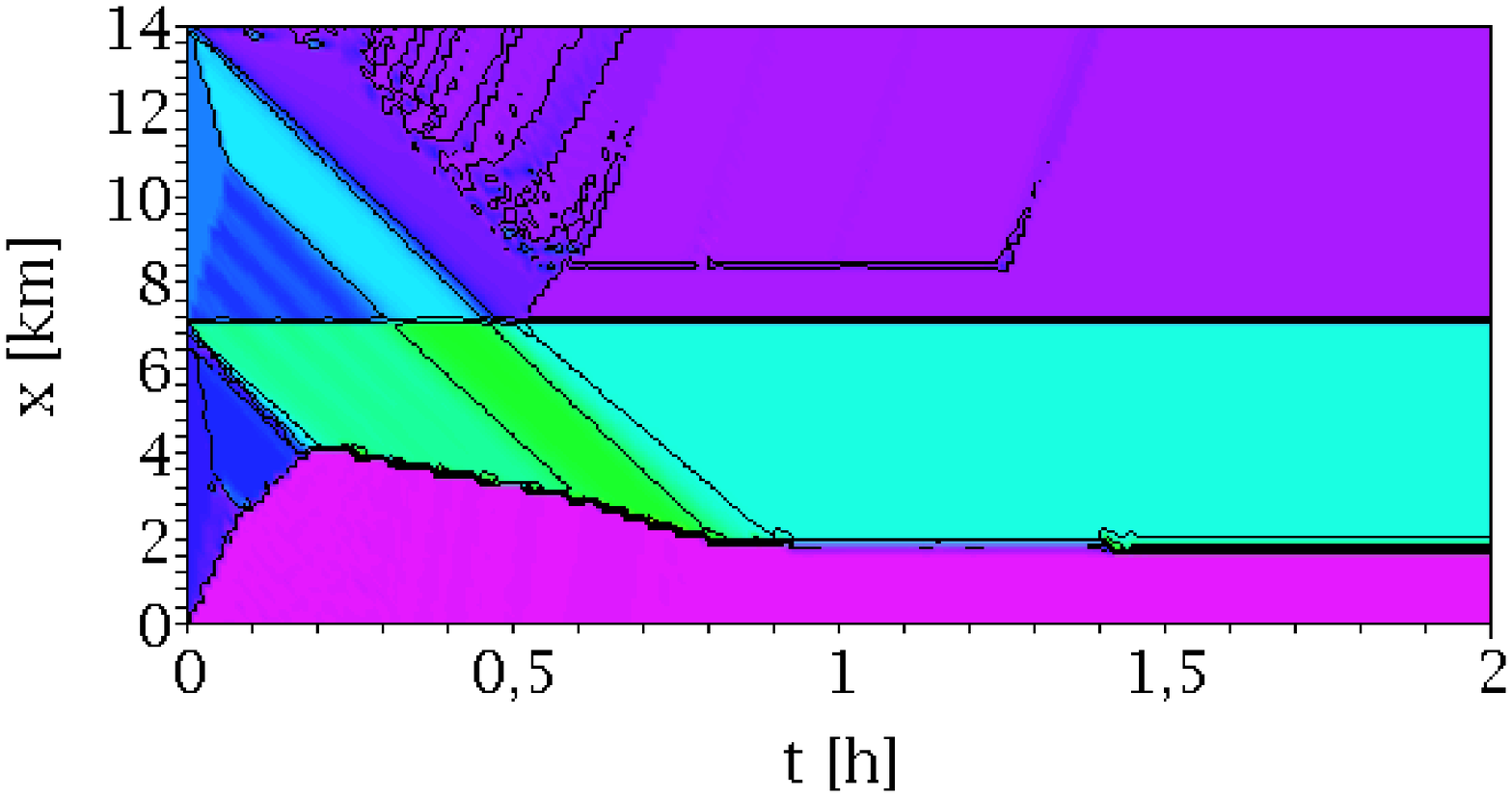,width=\linewidth} 
\end{minipage} \hfill
\begin{minipage}[b]{0.08\linewidth}
\centering \tiny $\rho_0=100$ [1/km]
\vspace{1.3cm}
\end{minipage} \hfill
\begin{minipage}[t]{0.45\linewidth}
 \centering\epsfig{figure=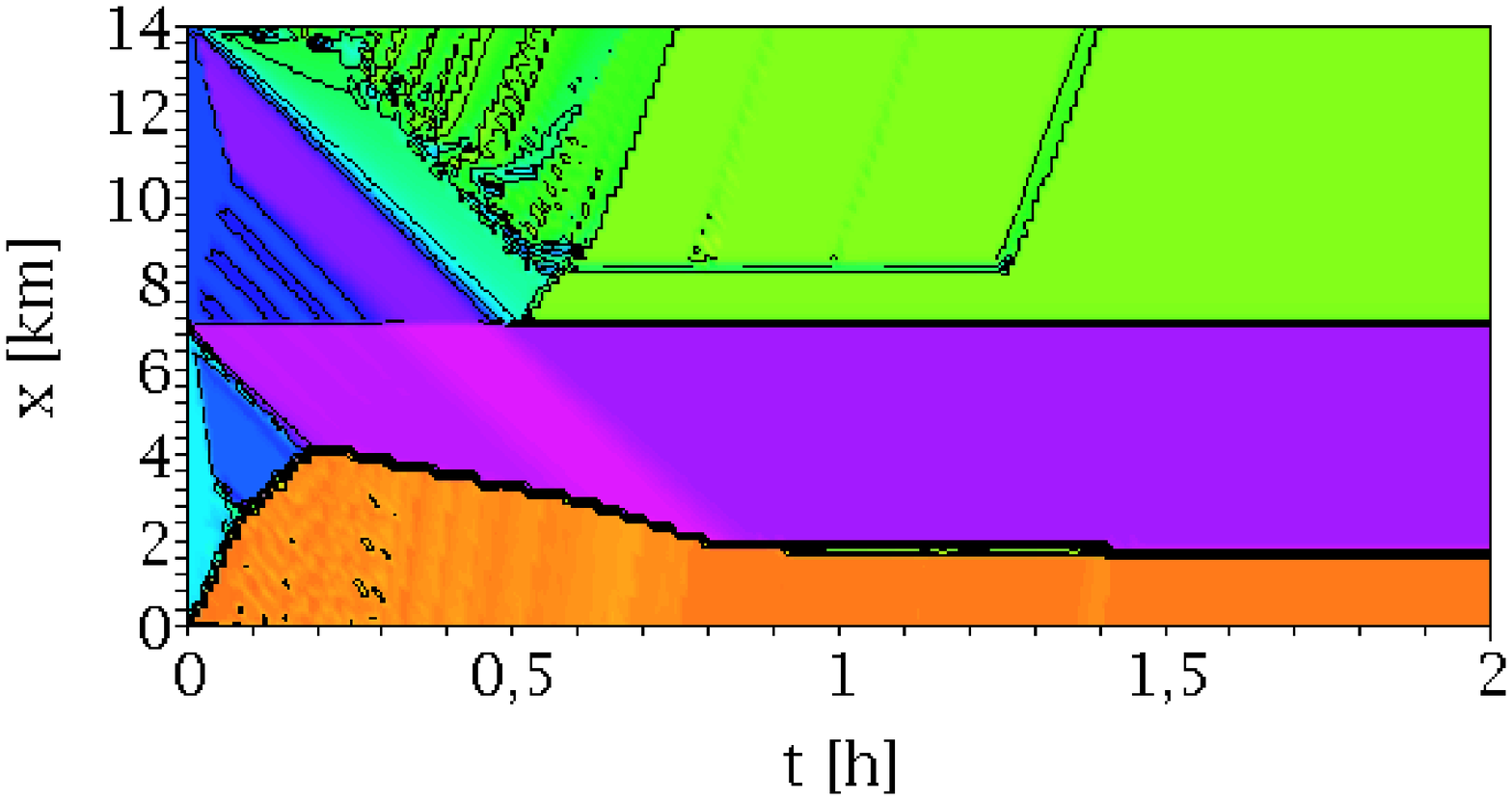,width=\linewidth}
\end{minipage}
\vspace{-0.2cm}
\begin{minipage}[t]{0.45\linewidth}
  \centering\epsfig{figure=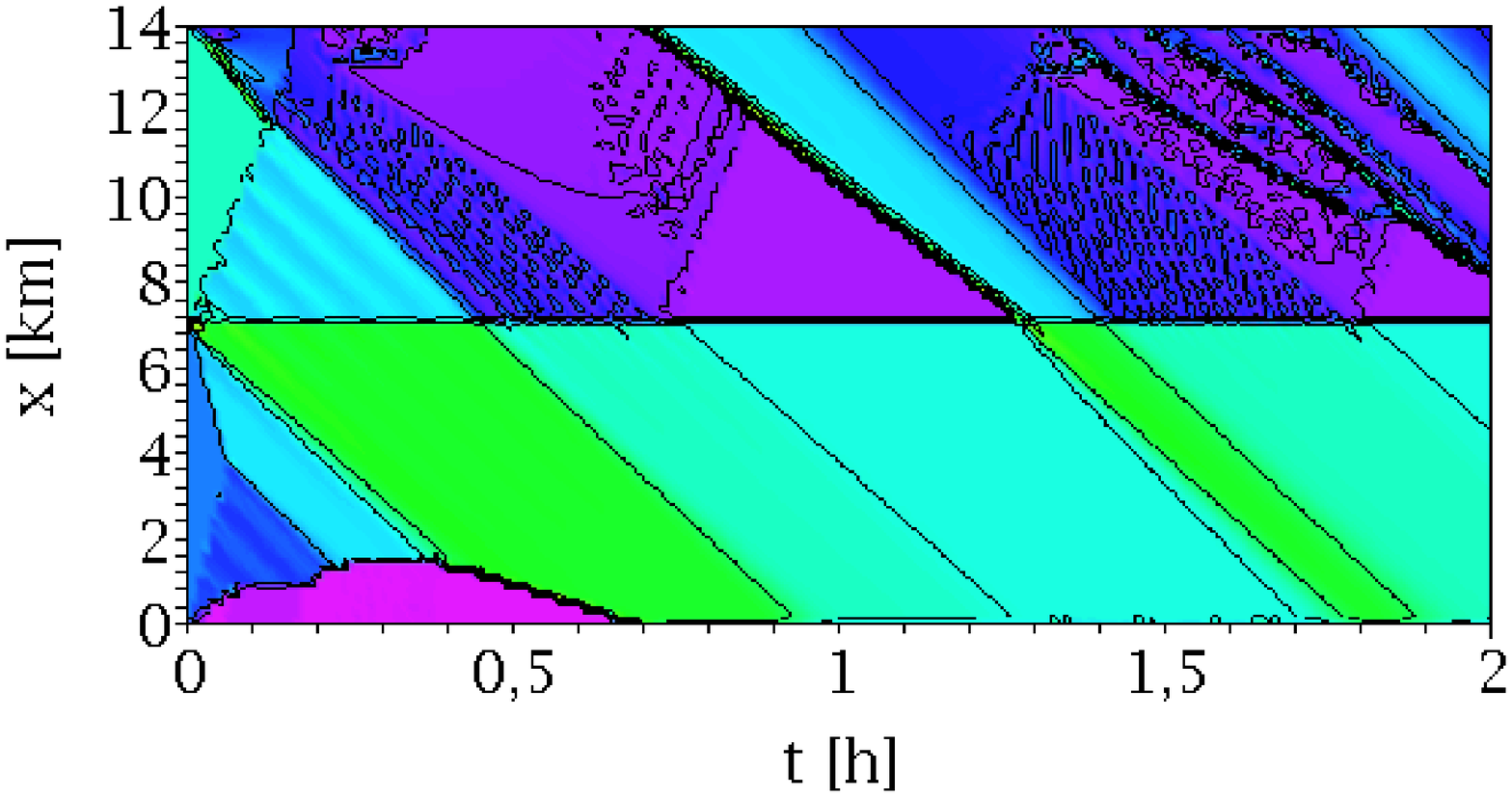,width=\linewidth} 
\end{minipage} \hfill
\begin{minipage}[b]{0.08\linewidth}
\centering \tiny $\rho_0=150$ [1/km]
\vspace{1.3cm}
\end{minipage} \hfill
\begin{minipage}[t]{0.45\linewidth}
 \centering\epsfig{figure=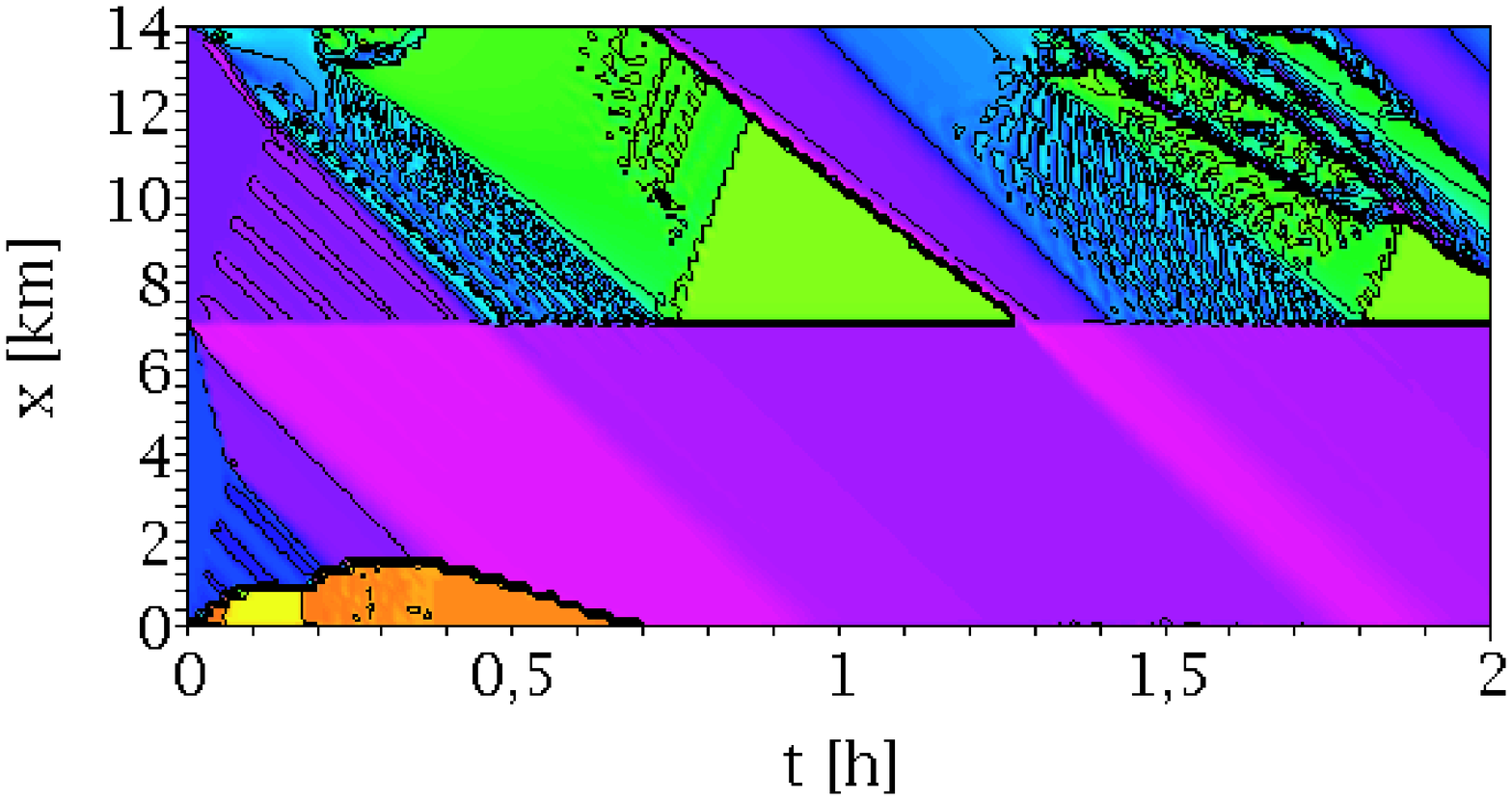,width=\linewidth}
\end{minipage}
\vspace{-0.2cm}
\begin{minipage}[t]{0.45\linewidth}
  \centering\epsfig{figure=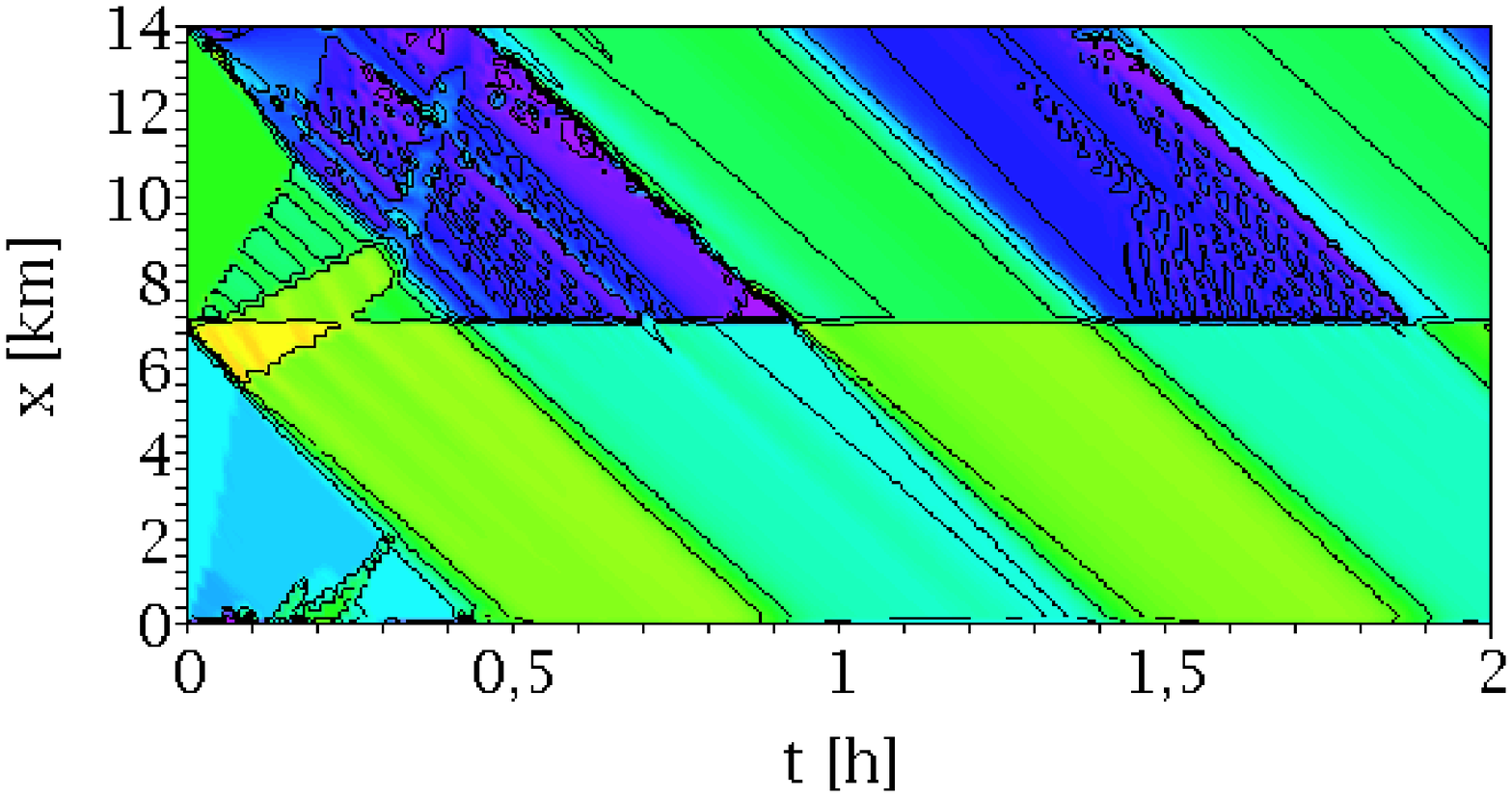,width=\linewidth} 
\end{minipage} \hfill
\begin{minipage}[b]{0.08\linewidth}
\centering \tiny $\rho_0=200$ [1/km]
\vspace{1.3cm}
\end{minipage} \hfill
\begin{minipage}[t]{0.45\linewidth}
 \centering\epsfig{figure=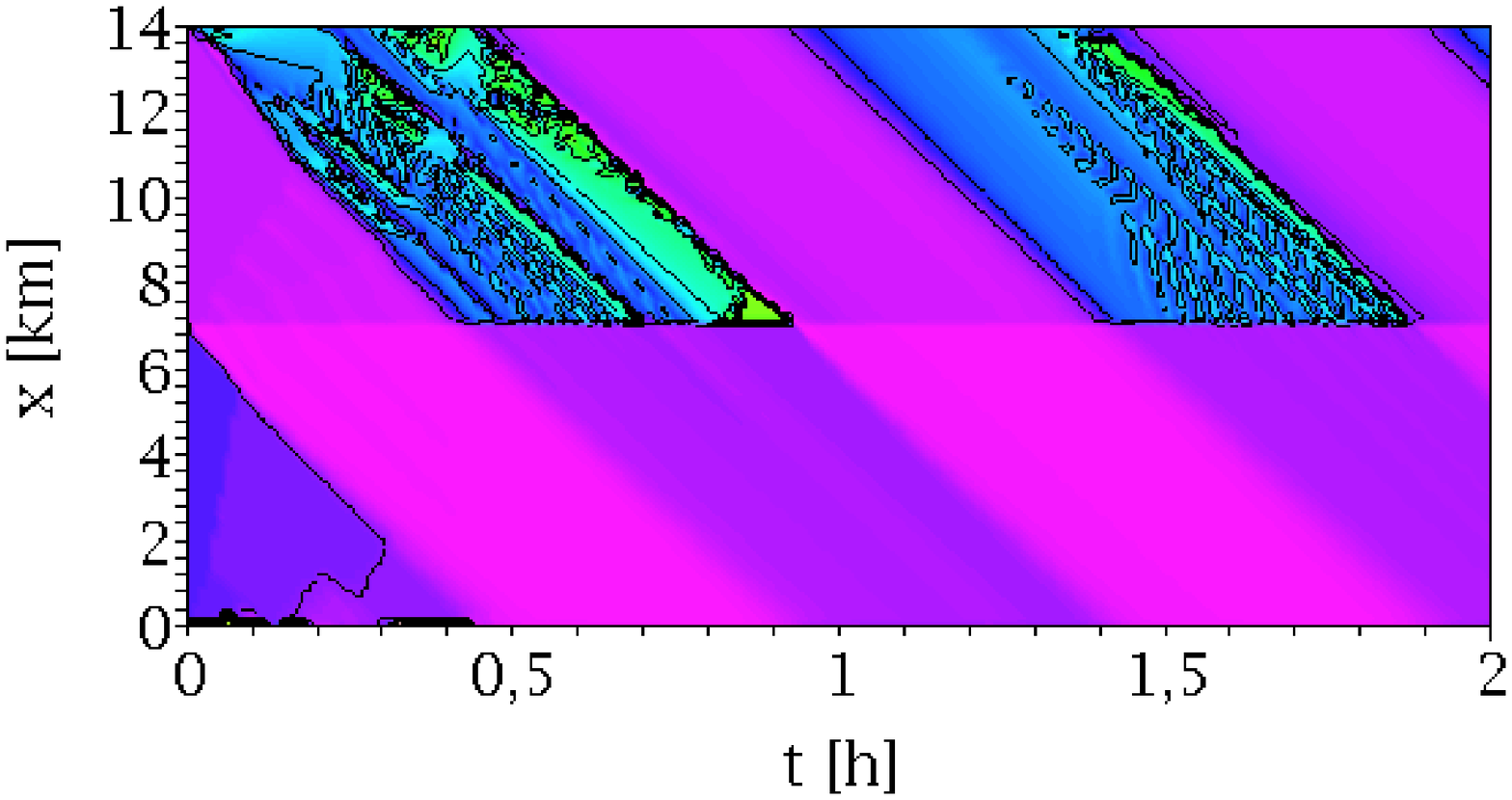,width=\linewidth}
\end{minipage}
\vspace{-0.2cm}
\begin{minipage}[t]{0.45\linewidth}
  \centering\epsfig{figure=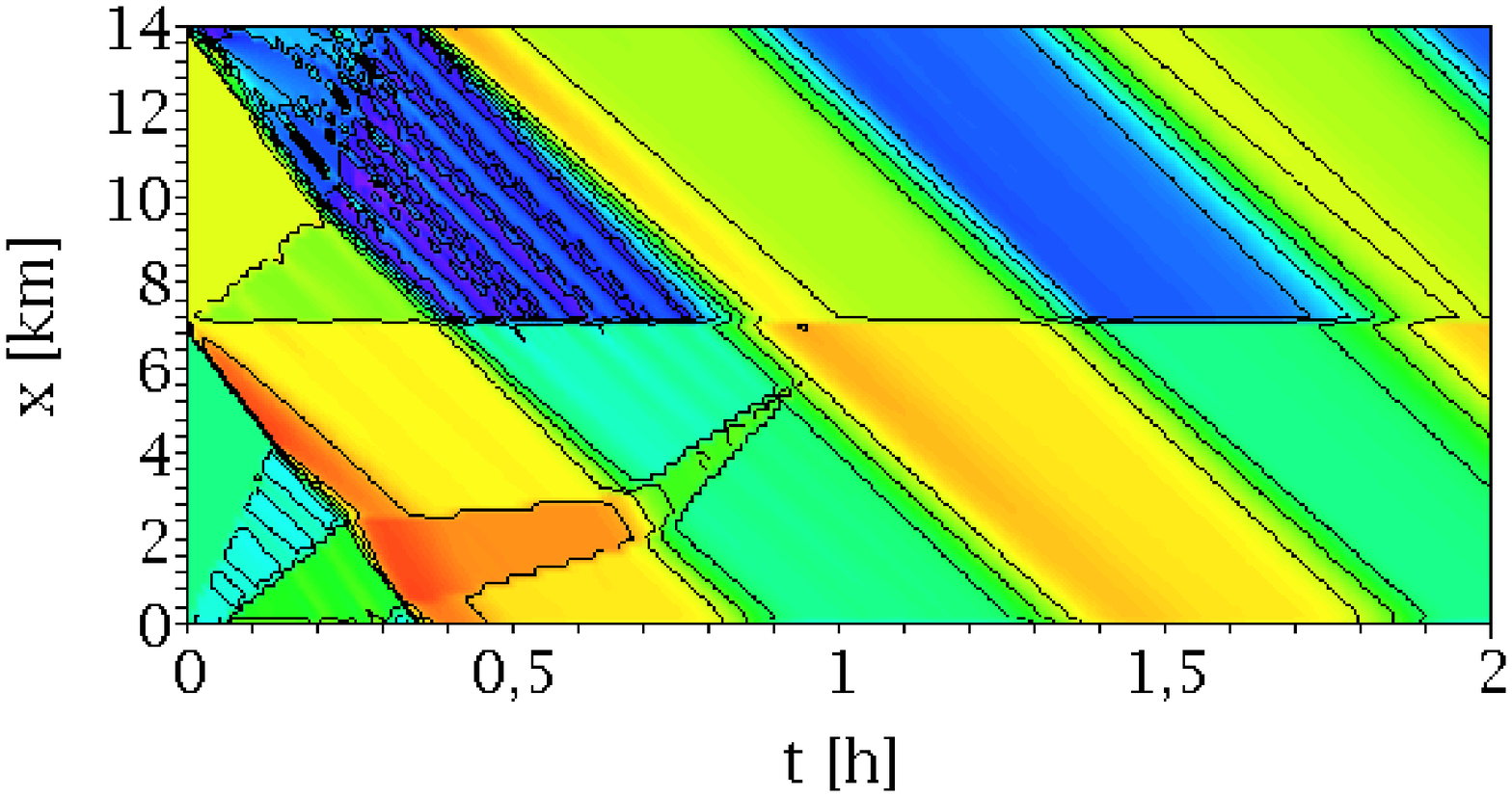,width=\linewidth} 
\end{minipage} \hfill
\begin{minipage}[b]{0.08\linewidth}
\centering \tiny $\rho_0=250$ [1/km]
\vspace{1.3cm}
\end{minipage} \hfill
\begin{minipage}[t]{0.45\linewidth}
 \centering\epsfig{figure=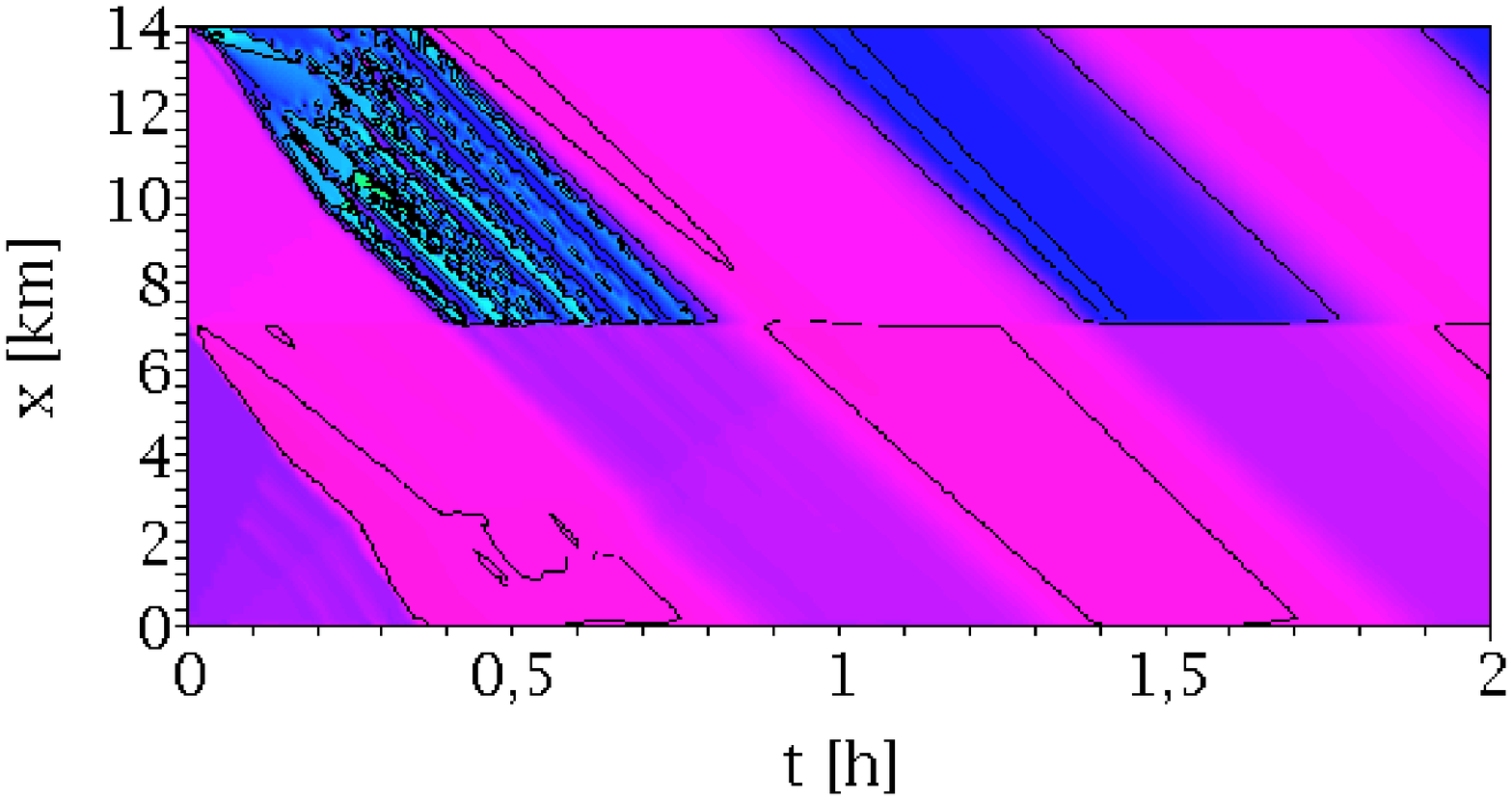,width=\linewidth}
\end{minipage}
\vspace{0cm}
\begin{minipage}[t]{0.45\linewidth}
  \centering\epsfig{figure=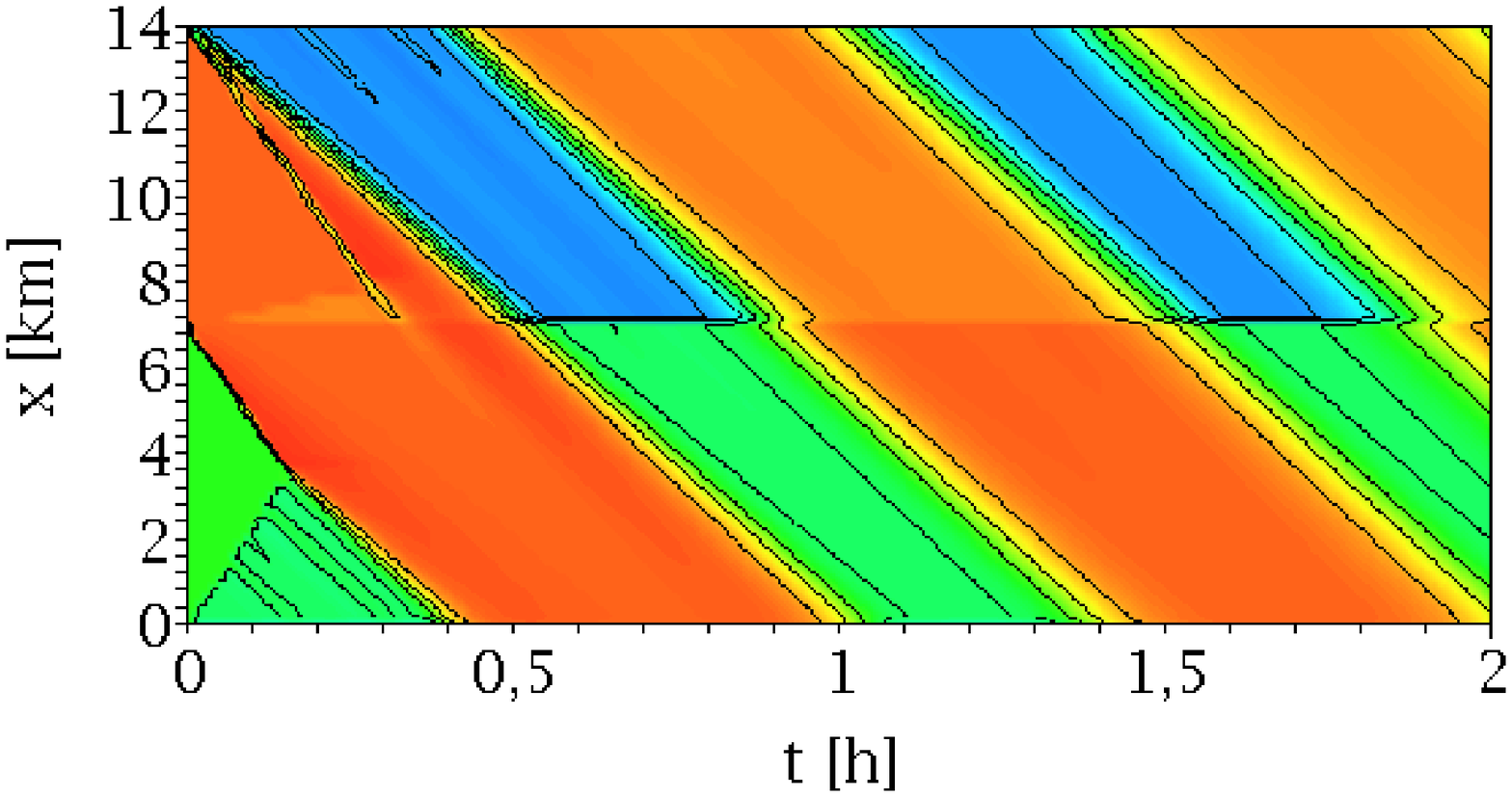,width=\linewidth} 
\end{minipage} \hfill
\begin{minipage}[b]{0.08\linewidth}
\centering \tiny $\rho_0=300$ [1/km]
\vspace{1.3cm}
\end{minipage} \hfill
\begin{minipage}[t]{0.45\linewidth}
 \centering\epsfig{figure=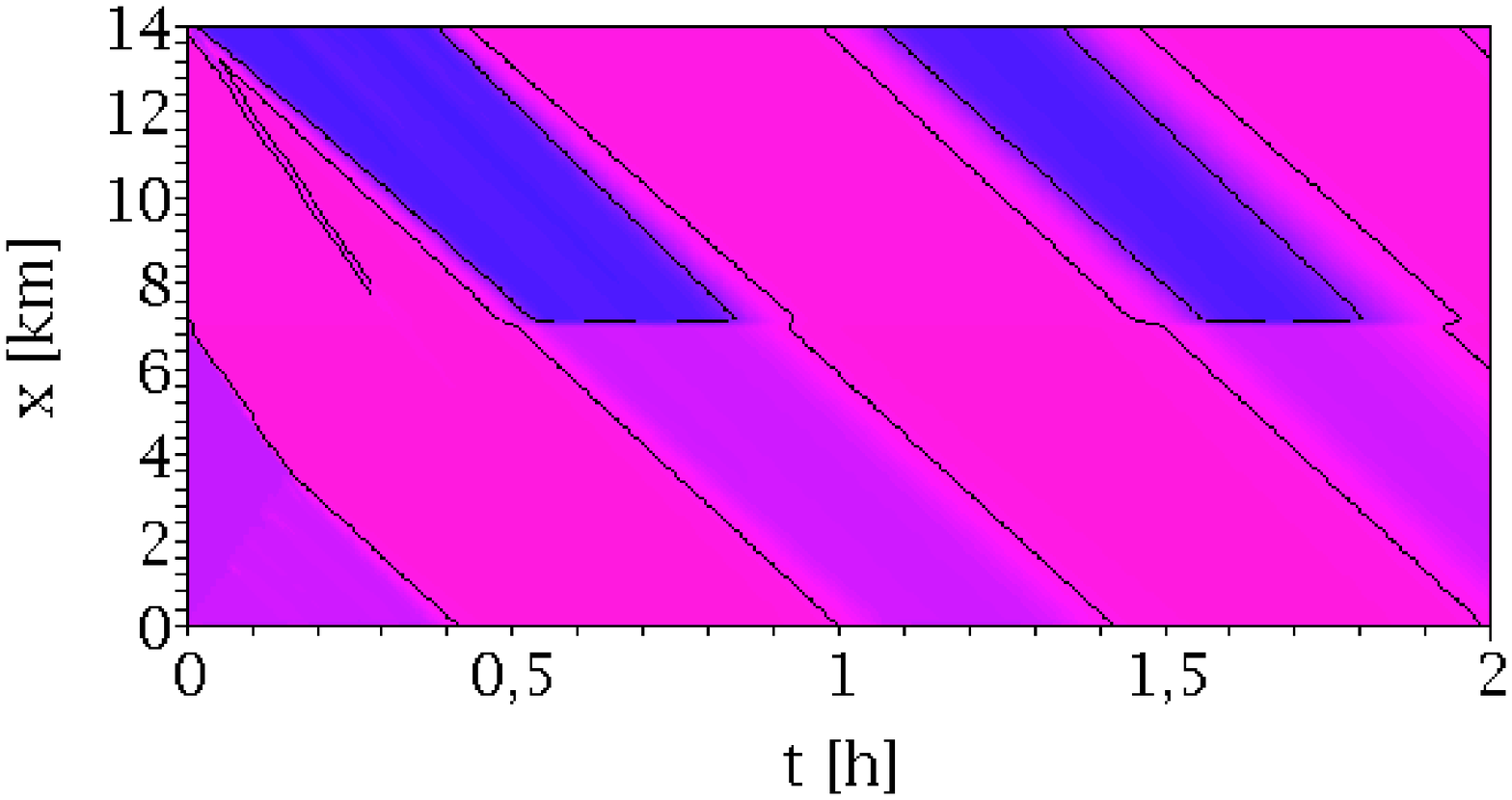,width=\linewidth}
\end{minipage}
\vspace{0cm}
\begin{minipage}[t]{0.45\linewidth}
  \centering\epsfig{figure=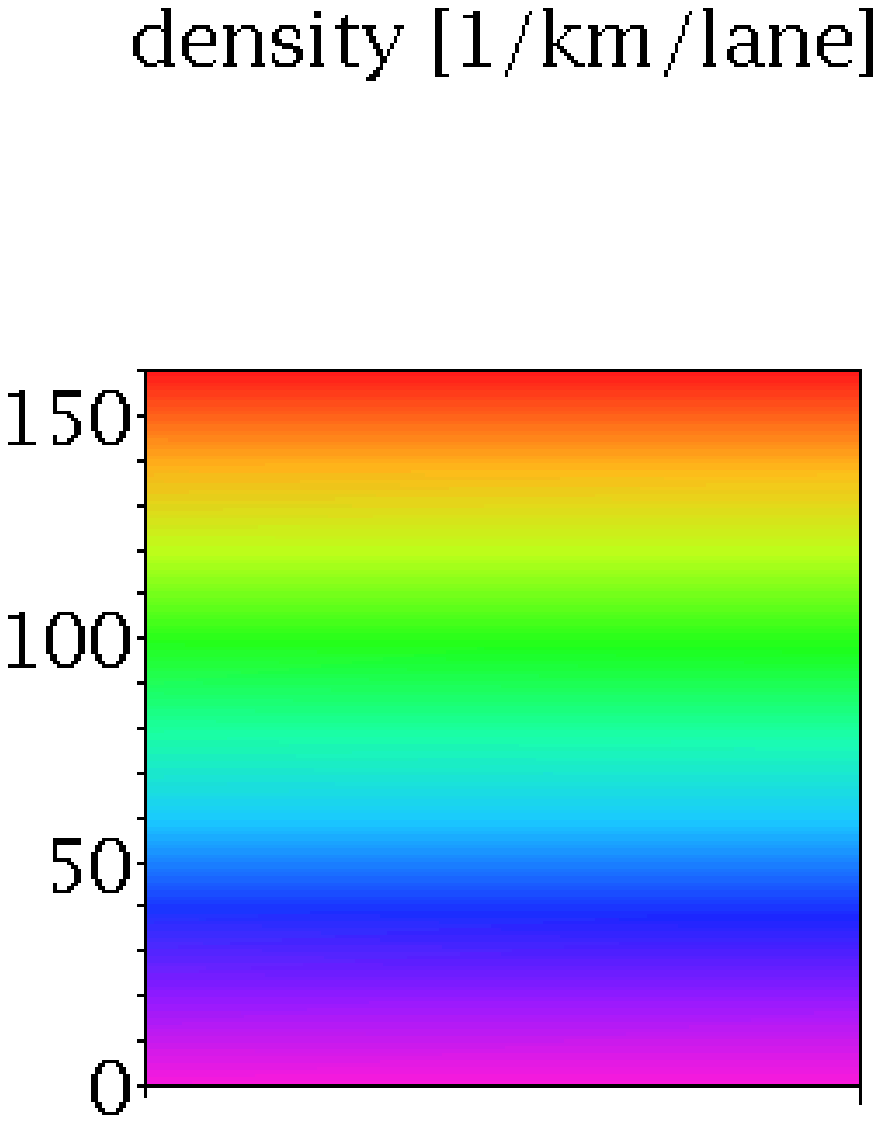,width=0.5\linewidth} 
\end{minipage} \hfill
\begin{minipage}[b]{0.08\linewidth}
\vspace{1.3cm}
\end{minipage} \hfill
\begin{minipage}[t]{0.45\linewidth}
 \centering\epsfig{figure=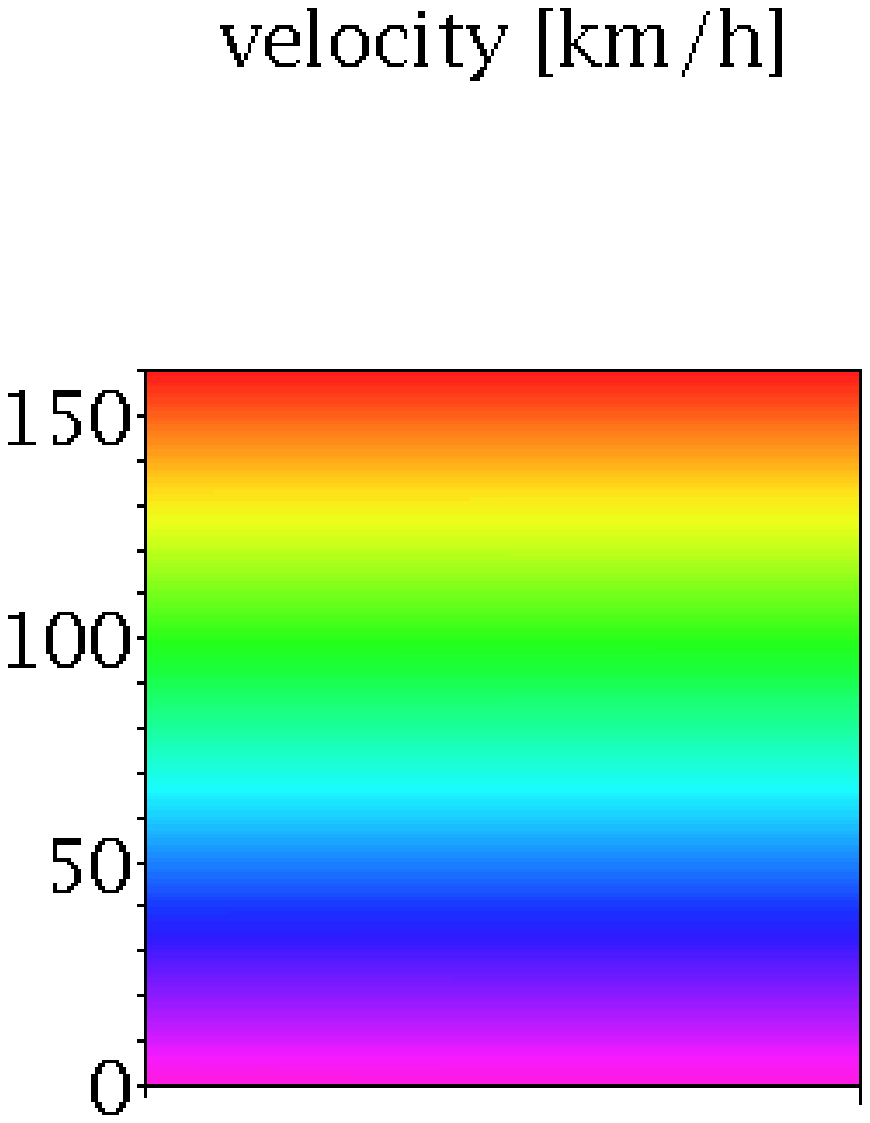,width=0.5\linewidth}
\end{minipage}
\caption{Traffic dynamics at the bottleneck caused by the reduction of the
  number of lanes on a highway. The column on the left shows the 
evolution
  of the vehicle density in units [1/km/lane], the column on the right the 
corresponding
  evolution of the velocity in units [km/h]. The different rows correspond to
  different simulation runs varying the initial density $\rho=\rho_0$ as
  indicated. See the text for a detailed description\label{fig3}.}
\end{figure}

We study the traffic dynamics for the setup depicted in Fig.~\ref{fig2}. The 
highway under study consists of two 7 km long road sections. The road section 
1 consists of three lanes whereas the road section 2 consists of two lanes. 
Note that in the mathematical description the transition from two to three 
lane is immediate, the length of the merging segments is neglected.
As in~\cite{SiM305} we use the equilibrium velocity function of Newell
\be
\label{u}
u(\rho) = u_m \Big(1- \exp \Big( -\frac{\lambda}{u_m}\Big(\frac{1}{\rho} -
\frac{1}{\rho_m} \Big) \Big) \Big)
\ee
with parameter values $u_m = 160 \ {\rm km/h}$, $\lambda = 3600 \ {\rm [1/h/lane]}$, $\rho_m = 160 \ {\rm [1/km/lane]}$ and an effective relaxation coefficient
\be
\label{beta}
b(\rho,v)=  \left\{
\begin{array}{ll}
\frac{a_c}{u-v},& \mbox{if~} \tilde{\beta}(\rho,v)(u(\rho)-v) - a_c \ge 0, \\
\frac{d_c}{u-v},& \mbox{if~} \tilde{\beta}(\rho,v)(u(\rho)-v) - d_c \le 0, \\
\tilde{\beta}(\rho,v),& \mbox{else},
\end{array} \right.
\ee
\be
\label{betatilde}
\tilde{\beta}(\rho,v) = \frac{1}{\hat{T} u_m} \Big( |u(\rho) - v + a_1 \Delta
v| + a_2 \Delta v \Big)
\ee
and
\be
\label{deltav}
\Delta v(\rho) = \tanh \Big( a_3 \frac{\rho}{\rho_m} \Big) \Big( u(\rho) + c
\rho_m \Big(\frac{1}{\rho} - \frac{1}{\rho_m}\Big) \Big),
\ee
with parameters $a_c = 2 \ {\rm m/s^2}$, $d_c = - 5 \ {\rm m/s^2}$, 
$\hat{T} = 0.1 \ {\rm s}$, $a_1 = - 0.2$, $a_2 = - 0.8$, $a_3 = 7$
and $c = -14 \ {\rm km/h}$. Thus the maximum density  of the road section 1 is 480
vehicles/km. The road section 2 can support 320 vehicles/km at maximum.
The initial data to start the numerical simulations consists of equilibrium
data on the two road sections. We prescribe a constant vehicle density
$\rho_0$ in both road sections, setting the initial velocity to $v =
u(\rho_0)$. We choose the constant $\rho_0$ to be independent of the number of
the lanes, the corresponding scaled densities in each road section follow from 
dividing $\rho_0$ by the number of lanes of that road section. In the 
following we perform a parameter study of $\rho_0$, varying the quantity 
between $50$ [1/km] and $300$ [1/km] in steps of $50$ [1/km]. 
Figure~\ref{fig3} displays the simulation results for the density 
(left column) and velocity 
(right column) for simulations covering two hours. Note that, although the
initial data are in equilibrium in each road section, the coupling conditions
at the interface between the two road sections do not guarantee the equilibrium
during the evolution.

For a density $\rho_0 = 50$ [1/km], a small region of higher density and lower
velocity forms between about 5.5 km and 7 km. This region
corresponds to data located in the fundamental diagram on and scattered around 
the jam  line. Clearly, this congested region is fixed at the bottleneck and 
therefore cannot correspond to a wide moving jam. The structure is supported 
by the bottleneck, i.e. the insufficient capacity of the road section 2 to 
carry the corresponding free flow rates in the road section 1. This region 
corresponds to synchronized flow.
For a density $\rho_0 = 100$ [1/km], the dynamics becomes more complicated,
but finally a synchronized flow region of extended width (ranging from about 
2 km to 7 km) forms. Only in a small region of the road section 1 (between 0 
km and 2 km) traffic is in free flow.
For a density $\rho_0 = 150$ [1/km], the synchronized flow region covers the
entire road section 1. Moreover, a wide moving jam travels through the two 
road sections,
see the bright pink structure in the velocity plot.
Increasing the density still further leads to even wider wide moving jams. 
The velocities inside these jams decrease with the increase of the initial 
density $\rho_0$. For $\rho_0 = 300$ [1/km] velocities of less than  1 [km/h] 
are reached inside the wide moving jam. Note that the wide moving jams are 
not affected by the interfaces between the two road sections of the highway. 
They travel upstream with an almost constant speed.
\section{Bottlenecks caused by on-ramps and off-ramps}
\label{onofframp}
\begin{figure}[htbp]
\centering\epsfig{figure=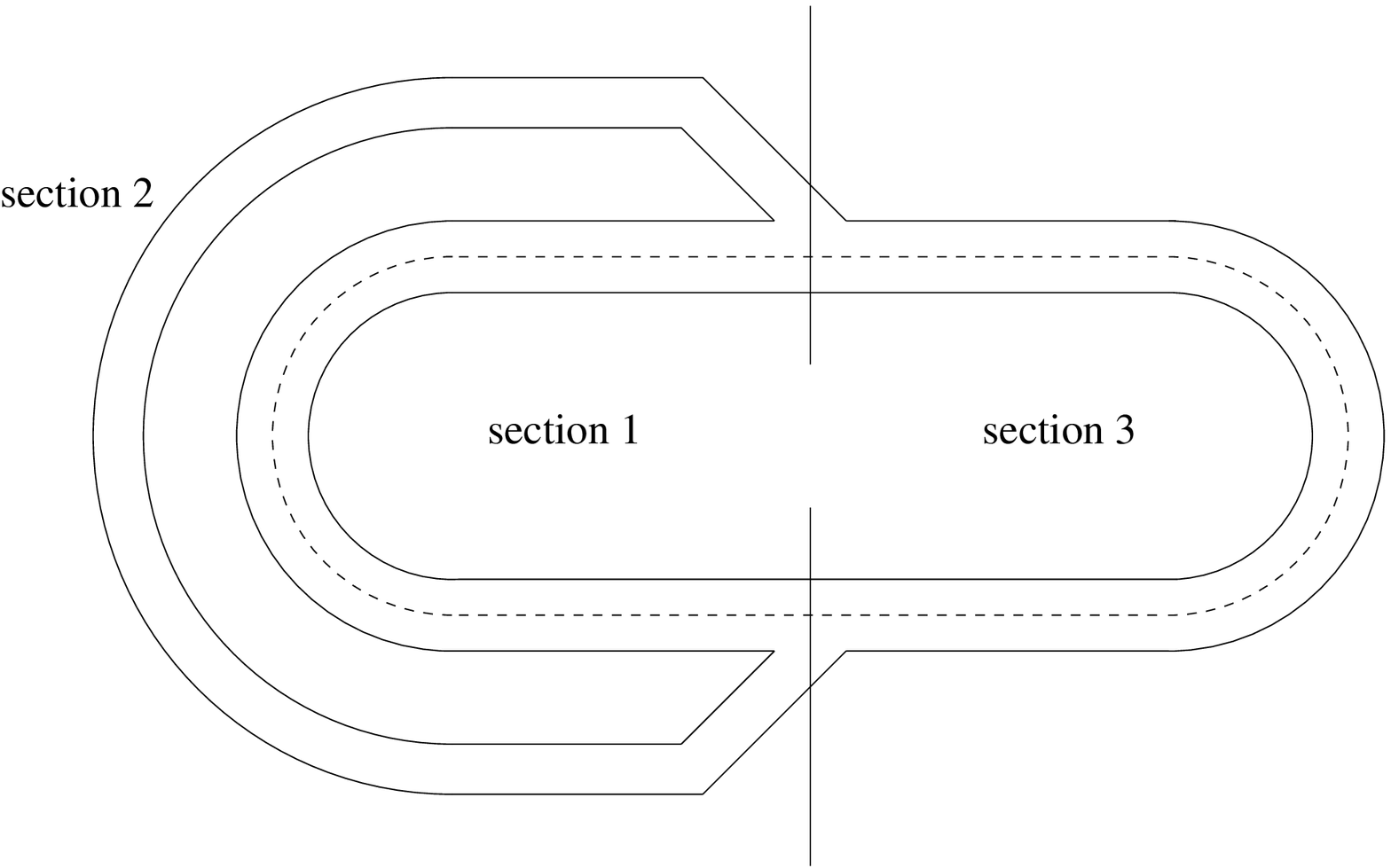,width=0.7\linewidth}
\caption{Sketch of the simulation setup. We study the dynamics on a two-lane 
highway with an on-ramp between the road sections 1 and 3 and an off-ramp
between the road sections 3 and 1. On- and off-ramp form the boundaries of the 
one-lane road section 2.\label{fig4}}
\end{figure}

\begin{figure}[htbp]
\vspace{-0.1cm}
\begin{minipage}[t]{.45\linewidth}
  \centering\epsfig{figure=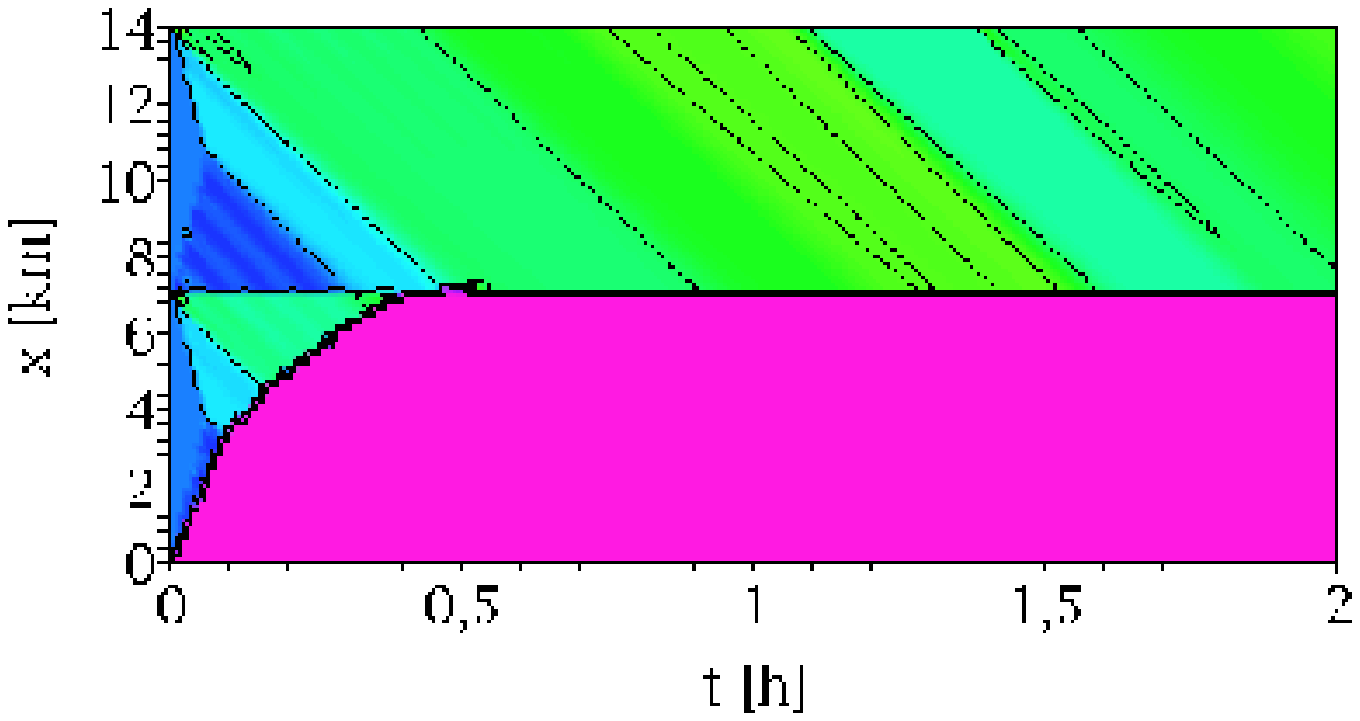,width=\linewidth} 
\end{minipage} \hfill
\begin{minipage}[b]{0.08\linewidth}
\centering \tiny $\alpha=0.1$
\vspace{1.3cm}
\end{minipage} \hfill
\begin{minipage}[t]{0.45\linewidth}
 \centering\epsfig{figure=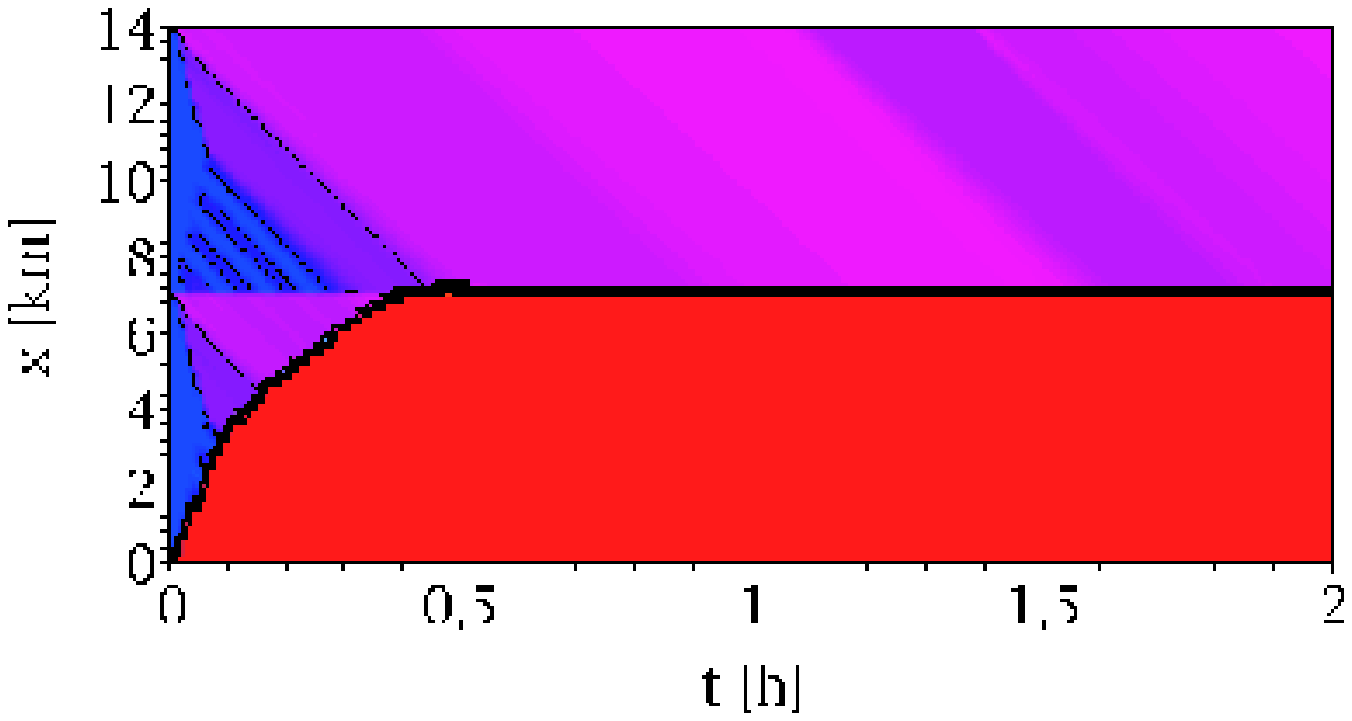,width=\linewidth}
\end{minipage}
\vspace{-0.2cm}
\begin{minipage}[t]{0.45\linewidth}
  \centering\epsfig{figure=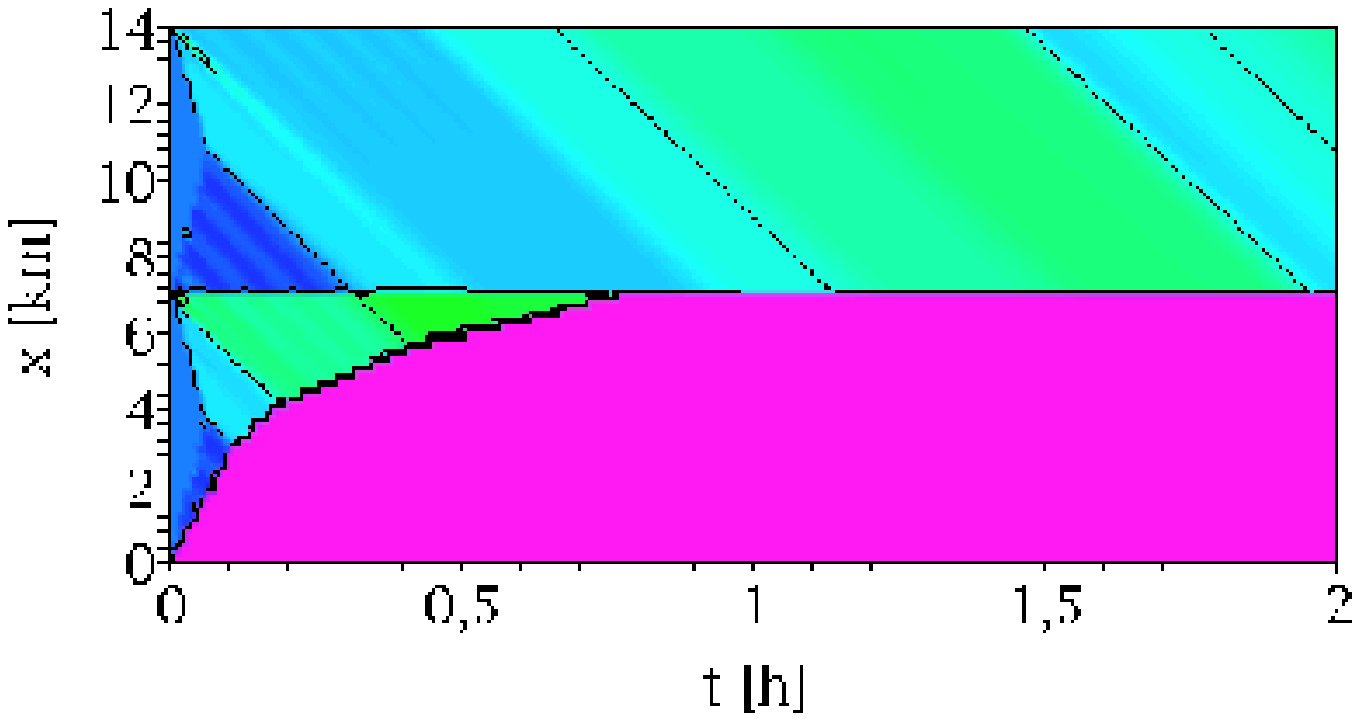,width=\linewidth} 
\end{minipage} \hfill
\begin{minipage}[b]{0.08\linewidth}
\centering \tiny $\alpha=0.3$
\vspace{1.3cm}
\end{minipage} \hfill
\begin{minipage}[t]{0.45\linewidth}
 \centering\epsfig{figure=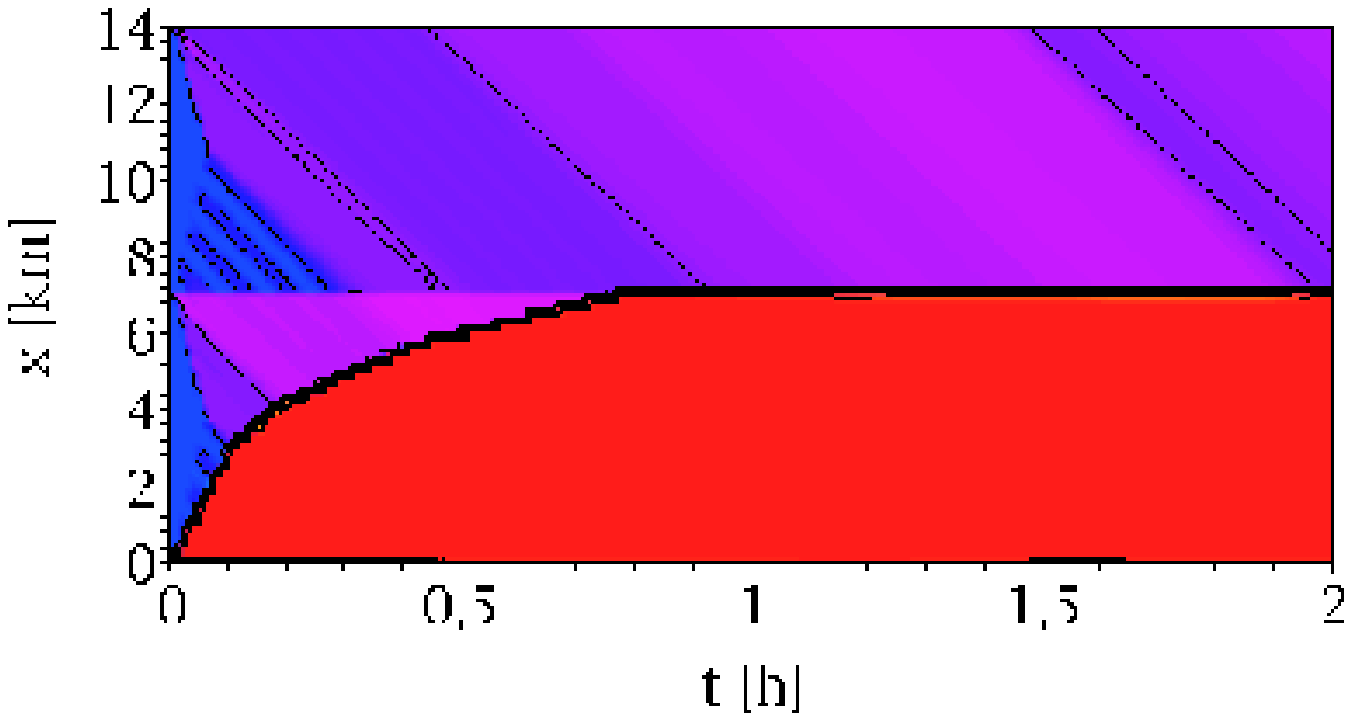,width=\linewidth}
\end{minipage}
\vspace{-0.2cm}
\begin{minipage}[t]{0.45\linewidth}
  \centering\epsfig{figure=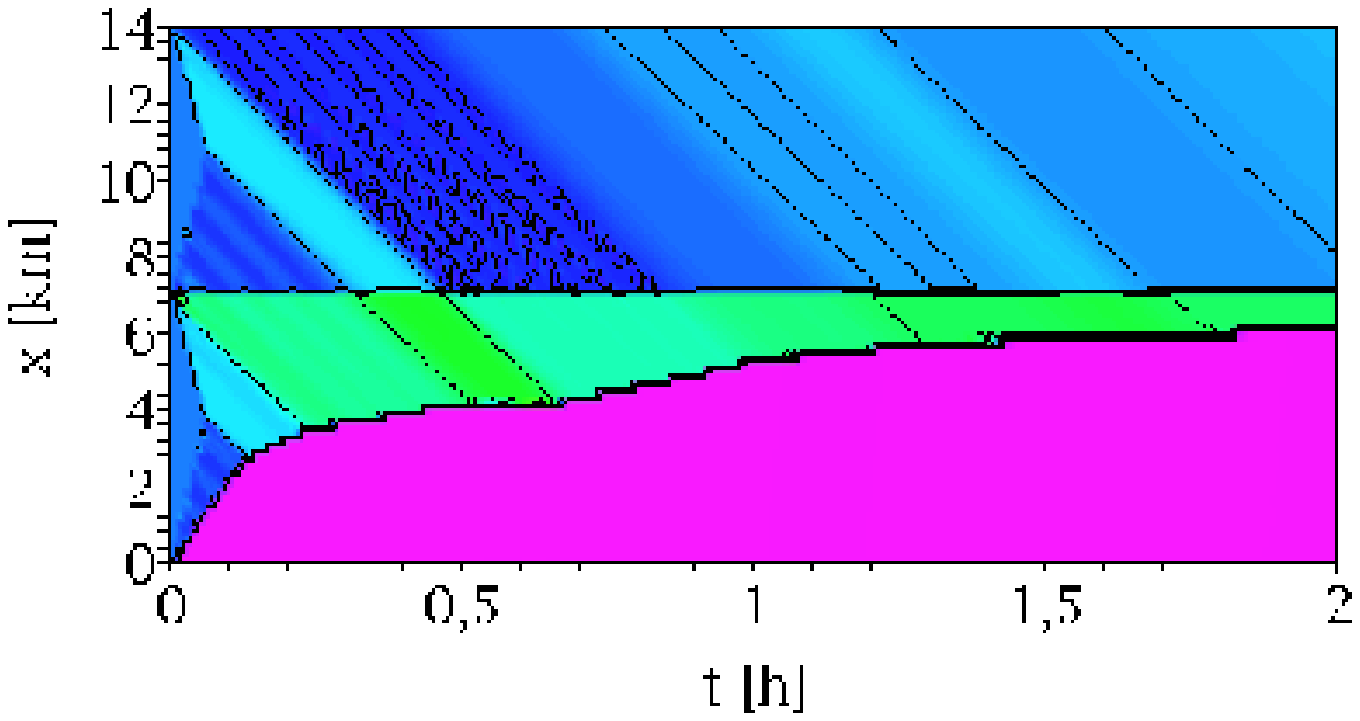,width=\linewidth} 
\end{minipage} \hfill
\begin{minipage}[b]{0.08\linewidth}
\centering \tiny $\alpha=0.5$
\vspace{1.3cm}
\end{minipage} \hfill
\begin{minipage}[t]{0.45\linewidth}
 \centering\epsfig{figure=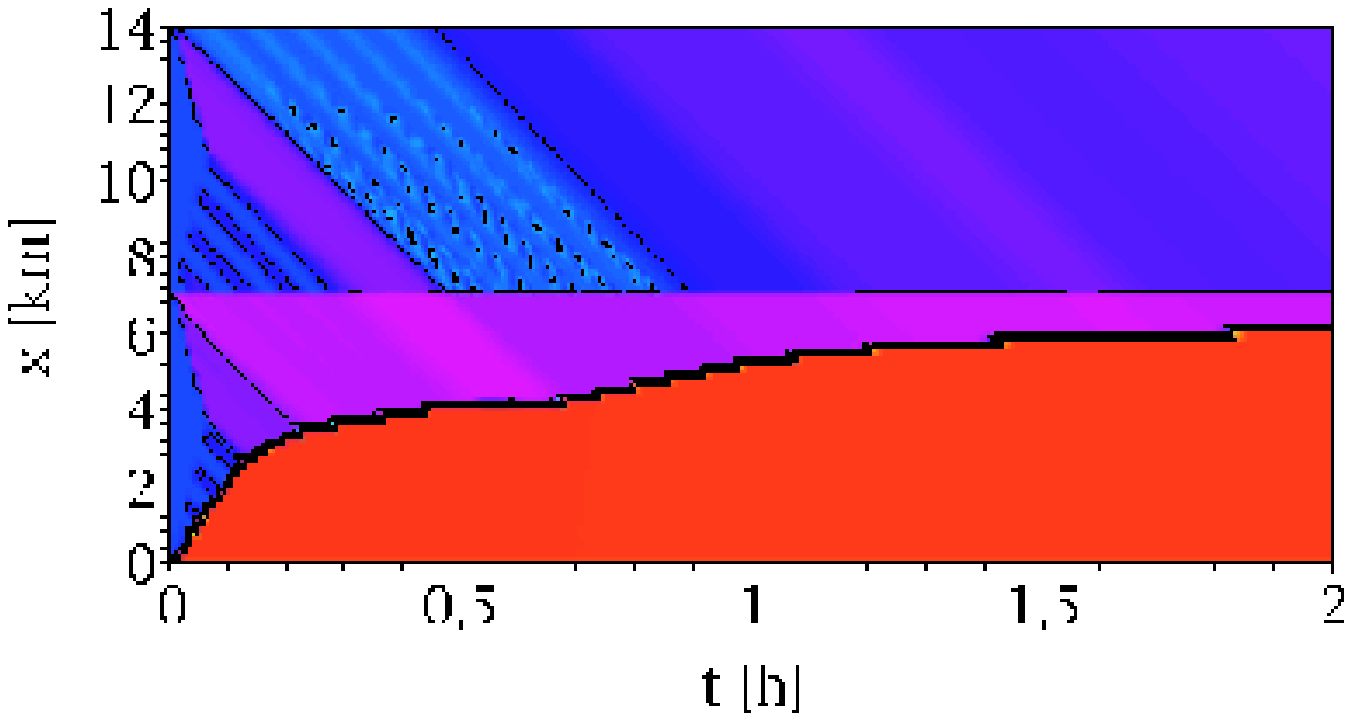,width=\linewidth}
\end{minipage}
\vspace{-0.2cm}
\begin{minipage}[t]{0.45\linewidth}
  \centering\epsfig{figure=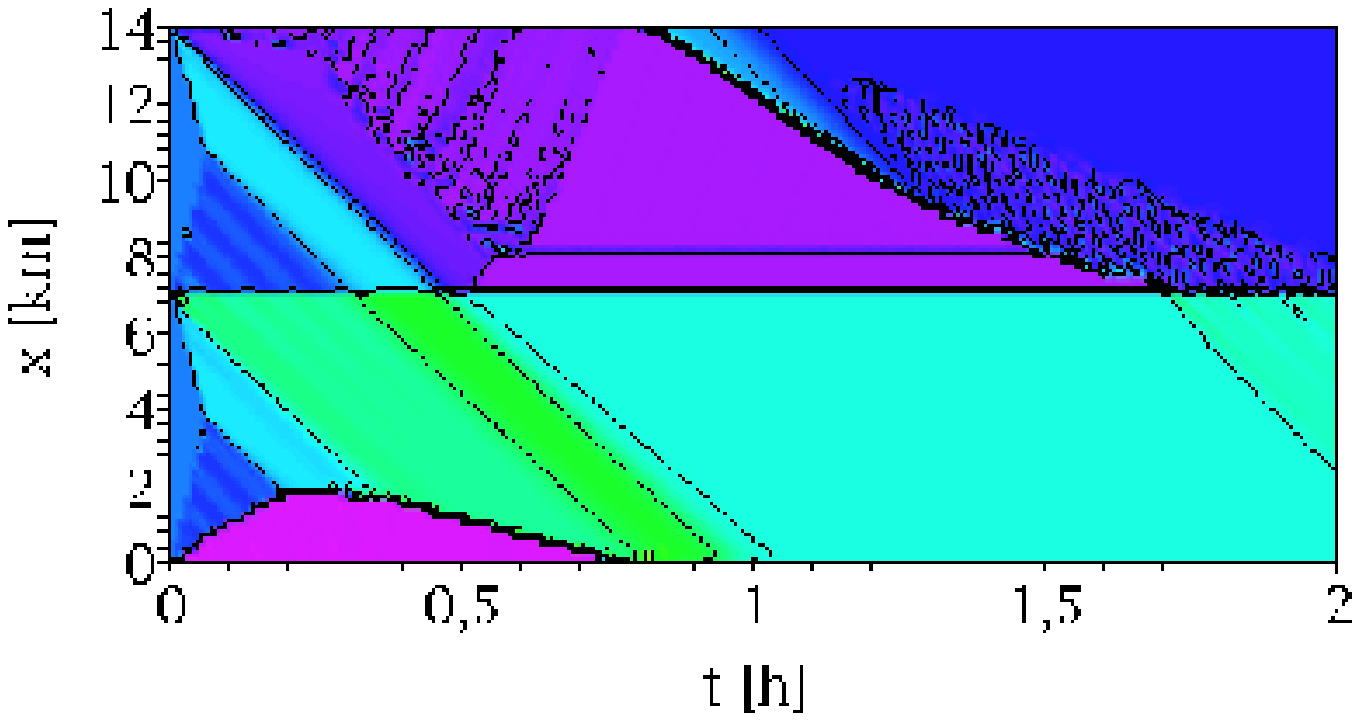,width=\linewidth} 
\end{minipage} \hfill
\begin{minipage}[b]{0.08\linewidth}
\centering \tiny $\alpha=0.7$
\vspace{1.3cm}
\end{minipage} \hfill
\begin{minipage}[t]{0.45\linewidth}
 \centering\epsfig{figure=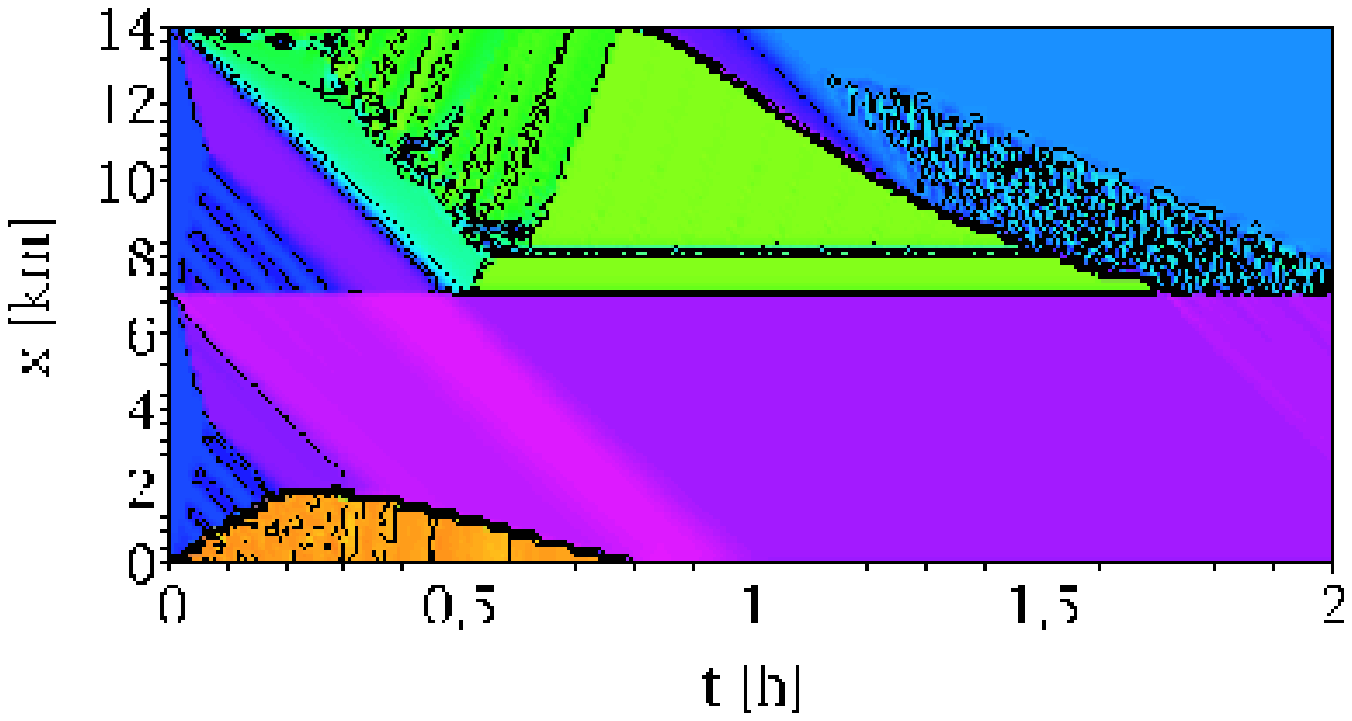,width=\linewidth}
\end{minipage}
\vspace{0cm}
\begin{minipage}[t]{0.45\linewidth}
  \centering\epsfig{figure=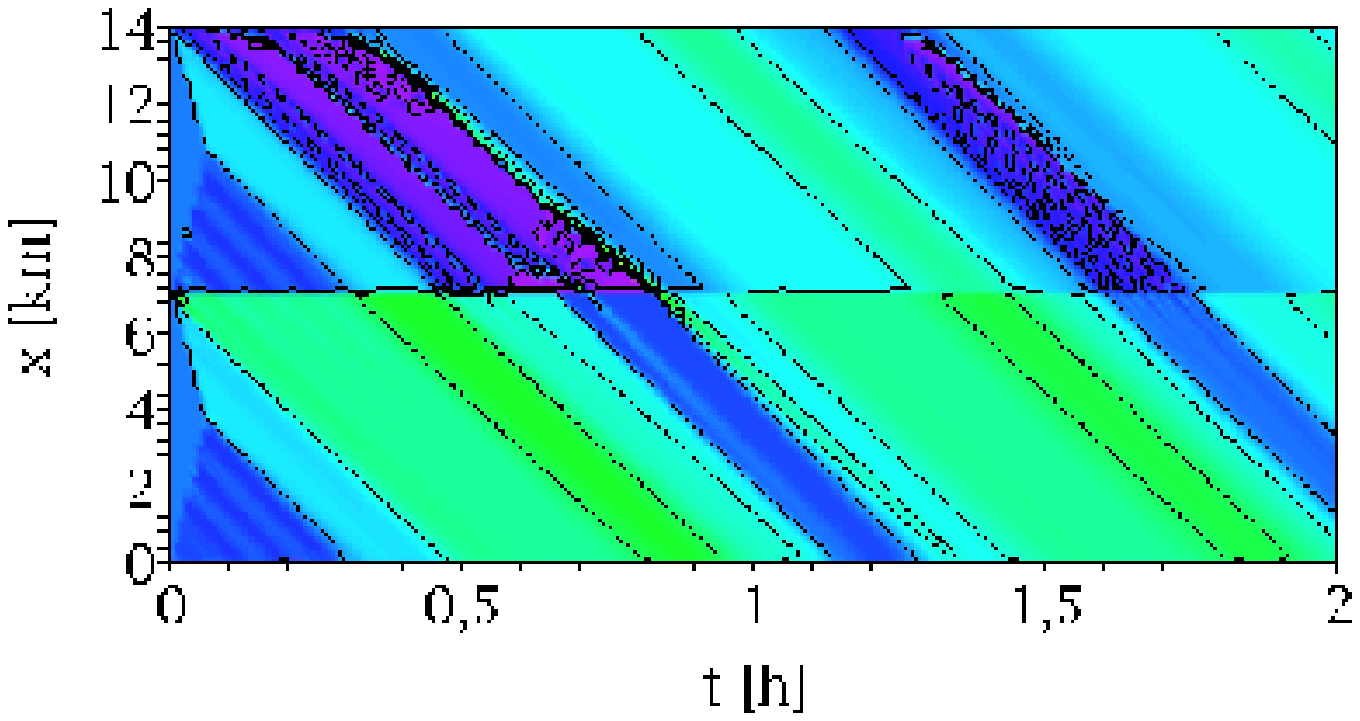,width=\linewidth} 
\end{minipage} \hfill
\begin{minipage}[b]{0.08\linewidth}
\centering \tiny $\alpha=0.9$
\vspace{1.3cm}
\end{minipage} \hfill
\begin{minipage}[t]{0.45\linewidth}
 \centering\epsfig{figure=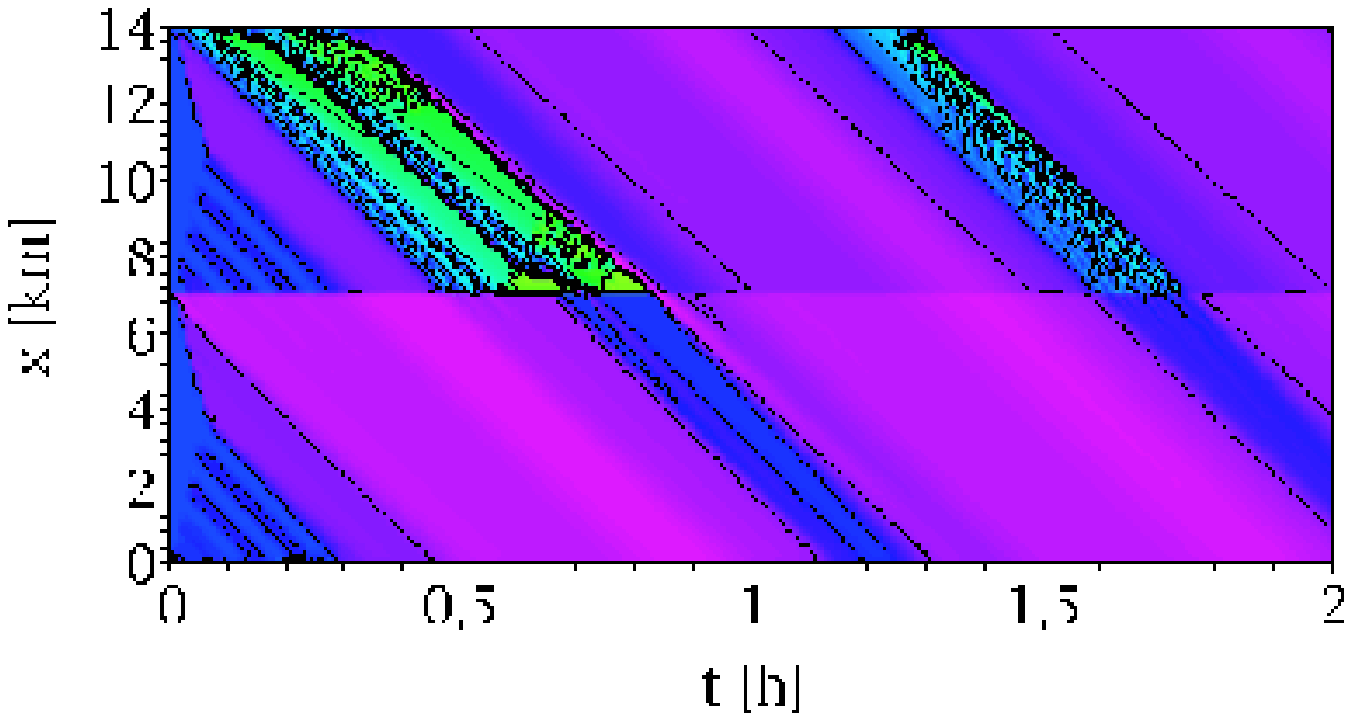,width=\linewidth}
\end{minipage}
\vspace{0cm}
\begin{minipage}[t]{0.45\linewidth}
  \centering\epsfig{figure=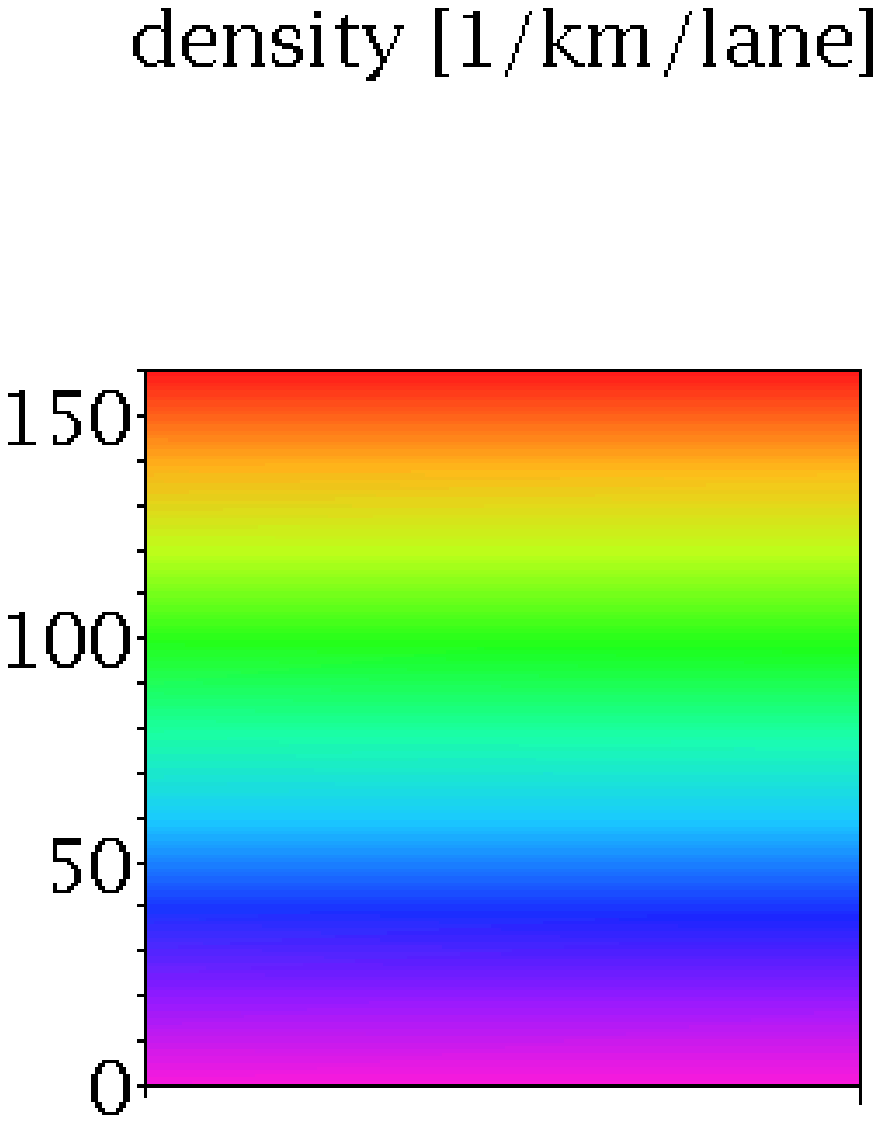,width=0.5\linewidth} 
\end{minipage} \hfill
\begin{minipage}[b]{0.08\linewidth}
\vspace{1.3cm}
\end{minipage} \hfill
\begin{minipage}[t]{0.45\linewidth}
 \centering\epsfig{figure=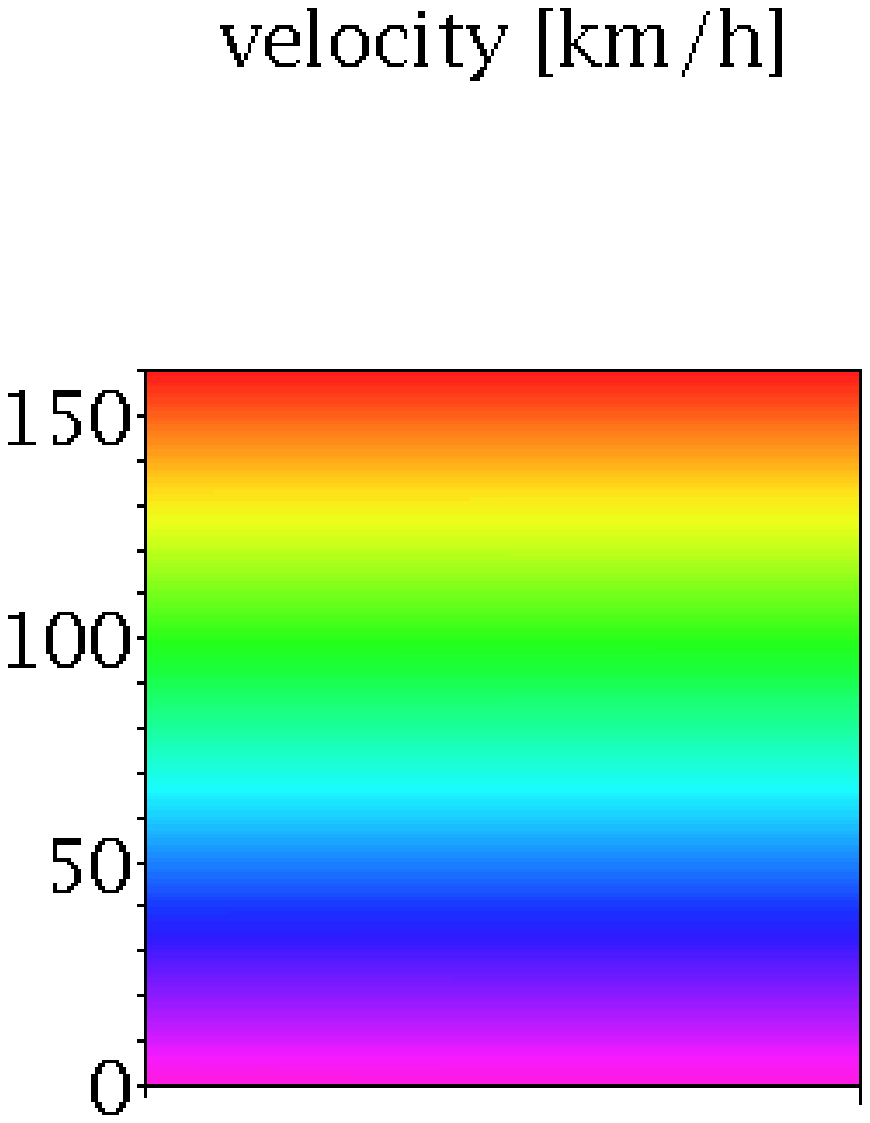,width=0.5\linewidth}
\end{minipage}
\caption{Traffic dynamics at highway bottlenecks caused by on-ramps and 
off-ramps. The plot shows the traffic dynamics on the two-lane highway,
the road section 1 corresponds to the region between 0 and 7 km, the road 
section 3 corresponds to the region between 7 and 14 km. We vary the
percentage of cars intending to enter from the road section 3 into the 
road section 1, i.e. the value of $\alpha$, in the range
between 0.1 and 0.9 (different rows). The column on the left shows the 
evolution  of the vehicle density in units [1/km/lane], the column on the
right the corresponding evolution of the velocity in units [km/h].\label{fig5}}
\end{figure}

In the second setup we analyze by numerical means a two-lane highway with an
on-ramp and an off-ramp. The simulation setup is displayed in Fig.~\ref{fig4}.
For our simulations we chose a length of 7 km for two-lane road sections 1 and 
3 each, and a length of 10 km for the one-lane road section 2. 
For the parameterization of the equilibrium velocity curve and the effective
relaxation coefficient, we use again the values given in
Eqs.~(\ref{u})-(\ref{deltav}). We start the simulations with a constant
  density in equilibrium on all road sections of $\rho_0 = 50$ [1/km/lane] and 
vary the percentage of cars $\alpha$ aiming to enter the road section 1 from
the road section 3, see Eqn.~(\ref{alpha}). The numerical results are
summarized in Fig.~\ref{fig5}.
For small to intermediate values of the parameter $\alpha$, the simulations
develop a state, where the road section 1 is almost empty, i.e. traffic is in 
free flow in the road section 1. This can be easily understood by realizing 
that for small
values of $\alpha$ most of the cars use the by-pass road section 2. In
contrast, traffic is in the congested regime in the road section 3. 
For sufficiently high values of $\alpha$ (see the results for $\alpha=0.5$)
synchronized flow develops in front of the on-ramp in the road section 1. 
For $\alpha=0.7$ there is only a small region of free flow remaining in road 
section 1 which disappears 
after a time of about 0.8 h. For a parameter value $\alpha=0.9$ a
region of narrow moving jams forms in front of the off-ramp in the road 
section 3.
\section{Extension to a general junction}
\label{extension}
For a given junction $n$, let us denote by  $\delta^-_n$ and $\delta^+_n$, 
respectively the set of all incoming roads  to $n$ (indexed $i$ in the sequel) 
and the set of all the outgoing roads to $n$ (indexed $k$ in the following).
We require the equations (\ref{rhob})-(\ref{vb}) to hold on each road of 
$\delta_n^-\cup \delta_n^+$.
The percentage of cars on the road $i$ intending to go to the road $k$ are denoted by 
$\alpha_{ik}$, such that $\forall i \in \delta_n^-, \sum_k \alpha_{ik}=1$. These 
coefficients are assumed to be known. 
 
Let $U_i^-=(\rho_i^-, \rho_i^- v_i^-), \ \ \forall i \in \delta_n^-$  and 
$U_k^-=(\rho_k^+, \rho_k^+ v_k^+), \ \ \forall k \in \delta_n^+$, respectively 
the boundary values on the incoming and outgoing roads. We denote by 
$\beta_{ik}$, such that $\forall k \in \delta_n^+, \sum_i \beta_{ik}=1$, the 
proportion of cars on the road $k$ coming from the road $i$. We set
\be 
\beta_{ik}=\frac{\alpha_{ik} d_i(\rho_i^-)}{\sum_{i \in \delta_n^-}
  {\alpha_{ik} d_i(\rho_i^-)}} \ \ \ \forall i \in \delta_n^-, \ \forall k \in \delta_n^+,
\ee
with the demand functions $d_i$ as defined in 
Eqs.~(\ref{demand1})-(\ref{demand2}). The homogenized $w$ on the outgoing 
roads near the junction are as follows 
\be
w_k^*=\sum_{i \in \delta_n^-} \beta_{ik} w_i (U_i^-), \ \ \ \forall k \in \delta_n^+.  
\ee
With these quantities we define the supply functions $s_k$ as in 
Eqs.~(\ref{supply1})-(\ref{supply2}) for arbitrary $k \in \delta_n^+$.
For all $k \in \delta_n^+$, the intermediate state of density $\rho_k^\dag$ on
the outgoing road is given by the intersection point between the curves 
$v_k(U)=v_k^+$ and $w_k(U)=w_k^*$.

To obtain the flux $q$ on each road one has to solve the following
maximization problem
\begin{subequations} \label{reformulate pb}
   \begin{eqnarray}
     \max \sum_{k \in \delta_n^+} q_k \mbox{ subject to }\\
      \ 0\leq q_i\leq d_i(\rho_i^-),    \forall i \in  \delta_n^-;\\
      \ 0\leq q_k\leq s_k(\rho_k^\dag),     \forall k \in \delta_n^+;\\
      \ q_i=\sum_{k \in \delta_n^+} \beta_{ik} q_k,    \forall i \in  \delta_n^-;\\
      \ q_k \leq \sum_{i \in \delta_n^-} \alpha_{ik} d_i(\rho_i^-),   \forall k \in \delta_n^+. 
  \end{eqnarray}
\end{subequations}
\section{Conclusion}
\label{conclusion}
We have studied the balanced vehicular traffic model at highway
bottlenecks caused by the reduction of the number of lanes and the effects of
an on-ramp and an off-ramp. To this
aim we performed numerical simulations changing the initial density and the
routing parameter at the off-ramp respectively. 
For the lane reduction setup the numerical results show that
already for moderate densities, the synchronized flow forms at the bottleneck. The
width of the synchronized flow region increases with increasing density. For 
large densities, wide moving jams appear which travel with a constant velocity
upstream. Wide moving jams are not affected by the interface between the 
highway sections, i.e. by a change in the number of lanes on the highway.
For the setup with an on-ramp and an off-ramp our numerical simulation show
that synchronized flow can form upstream of the on-ramp, but also in front of
an off-ramp, where narrow moving jams can emerge.

The theory of the coupling conditions described in this paper can be applied
to the balanced vehicular traffic model at a general junction, thus 
guaranteeing the conservation of the fluxes in the corresponding Riemann 
problems at intersections.
\section*{Acknowledgments}
F. Siebel would like to thank the Laboratoire J. A. Dieudonn\'e at Univerist\'e
Nice Sophia-Antipolis for the support and hospitality.
\\This work has been partially supported by the French ACI-NIM (Nouvelles 
Interactions des Math\'ematiques) N$^o$ 193 (2004).


\begin{thebibliography}{10}
\expandafter\ifx\csname url\endcsname\relax
  \def\url#1{\texttt{#1}}\fi
\expandafter\ifx\csname urlprefix\endcsname\relax\def\urlprefix{URL }\fi

\bibitem{AwR00}
A.~{Aw}, M.~{Rascle}, {Resurrection of ``second order'' models of traffic flow},
  SIAM Journal on Applied Mathematics 60 (2000) 916--938.

\bibitem{GaP06}
M.~{Garavello}, B.~{Piccoli}, {Traffic flow on a road network using the
  Aw-Rascle model}, to appear in Comm. Partial Differential Equations.

\bibitem{Gre01}
J.~{Greenberg}, {Extensions and amplifications of a traffic model of Aw and
  Rascle}, SIAM Journal on Applied Mathematics 62 (2001) 729--745.

\bibitem{Gre04}
J.~{Greenberg}, {Congestion Redux}, SIAM Journal on Applied Mathematics 64, (2004) 1175--1185.

\bibitem{GKR}
J.~{Greenberg}, A.~{Klar}, M.~{Rascle}, {Congestion on Multilane Highways}, 
SIAM Journal on Applied Mathematics  63 (2002) 818--833.

\bibitem{HaB05}
B.~{Haut}, G.~{Bastin}, {A second order model for road traffic networks}, in:
  {Proceedings of the 8th International IEEE Conference on Intelligent
  Transportation Systems}, (2005) 178--184.

\bibitem{HMR06}
M.~{Herty}, S.~{Moutari}, M.~{Rascle}, {Optimization criteria for modelling
  intersections of vehicular traffic flow}, Networks and Heterogenous Media 1
  (2006) 275--294.

\bibitem{HeR06}
M.~{Herty}, M.~{Rascle}, {Coupling conditions for a class of "second-order"
  models for traffic flow}, SIAM Journal on Mathematical Analysis 38 (2006) 
  595--616.

\bibitem{Ker04}
B.~{Kerner}, {The Physics of Traffic}, Springer, Berlin, 2004.

\bibitem{SiM205}
F.~{Siebel}, W.~{Mauser}, {On the fundamental diagram of traffic flow}, SIAM
  Journal on Applied Mathematics 66 (2006) 1150--1162.

\bibitem{SiM305}
F.~{Siebel}, W.~{Mauser}, {Synchronized flow and wide moving jams from balanced
  vehicular traffic}, \pre 73~(6) (2006) 066108.

\bibitem{SiM05}
F.~{Siebel}, W.~{Mauser}, {Simulating vehicular traffic in a network using
  dynamic routing}, Mathematical and Computer Modelling of Dynamical Systems,
  {\it in press}.

\bibitem{Zha02}
H.~M. {Zhang}, {A non-equilibrium traffic model devoid of gas-like behaviour},
  Transportation Research B 36 (2002) 275--290.

\end{thebibliography}
\end{document}